\documentclass{aa} 
\usepackage{graphicx} 
\usepackage{epsfig} 
\usepackage{amsmath} 
\usepackage[round]{natbib} 
\usepackage{hyperref}
\hypersetup{pdfauthor=Kai-Martin Dittkrist}
\hypersetup{backref=true, pagebackref=true, hyperindex=true, breaklinks=true,colorlinks=true,urlcolor=blue, linkcolor=blue,  citecolor=blue,pagecolor=red, bookmarks=true, bookmarksopen=true}
 
\usepackage{txfonts}

\def\mearth{\rm M_\oplus} 
\def\msun{\rm M_\odot} 
\def\rhill{R_{\rm H}}

\def\cs{c_{\rm s}} 
\def\fpg{f_{\rm D/G}}

\def\f1{f_{\rm I}} 
\def\mstar{M_*}

\def\mdotcore{\dot{M}_{\rm core}} 
 
\def\mwind{\dot{M}_{\rm w}} 
 
\def\miso{M_{\rm iso}}

\def\beq{\begin{equation}} 
\def\eeq{\end{equation}} 
 
\def\mplanet{M_{\rm p}} 
\def\aplanet{a_{\rm p}}

\def\cadia{C_{\rm adia }} 
\def\cloc{C_{\rm loc }} 
\def\clind{C_{\rm Lind}} 
\def\chsa{C_{\rm HS,adia}} 
\def\chsl{C_{\rm HS,loc}} 
\def\fvisc{f_{\rm visc }} 
\def\fcool{f_{\rm cool }} 
\def\fsigma{f_{\rm \Sigma}} 
\def\xs{x_{\rm s}} 
\def\lcool{l_{\rm cool}} 
\def\rp{a_{\rm p}} 
\def\omegap{\Omega_{\rm p}} 
\def\sigmap{\Sigma_{\rm p}} 
\def\tauuturn{\tau_{\rm u-turn}} 
\def\taucool{\tau_{\rm cool}} 
\def\taulib{\tau_{\rm lib}} 
\def\tauvisc{\tau_{\rm visc}} 
\def\msat{M_{\rm sat}} 
\def\mgap{M_{\rm gap}} 
 
\def\betat{\beta_{\rm T}} 
\def\betas{\beta_{\rm \Sigma}}

\def\({\left(} 
\def\){\right)} 
\def\<{\left<} 
\def\>{\right>}

\begin{document}

\title{Impacts of planet migration models on planetary populations}  
\subtitle{Effects of saturation, cooling and stellar irradiation}

\author{K.-M. Dittkrist\inst{1} \and C. Mordasini\inst{1}\thanks{Reimar-L\"ust
    Fellow of the MPG} \and H. Klahr\inst{1} \and Y. Alibert\inst{2,3}  \and 
  T. Henning\inst{1}}

\institute{ Max-Planck-Institut f\"ur Astronomie, K\"onigstuhl 17, D-69117 Heidelberg, Germany \and 
Physikalisches Institut, University of Bern, Sidlerstrasse 5, CH-3012 Bern, Switzerland \and   
Institut UTINAM, CNRS-UMR 6213, Observatoire de Besan\c{c}on, BP 1615, 25010 Besan\c{c}on Cedex, France }

\offprints{Kai-Martin DITTKRIST, \email{dittkrist@mpia.de}}

\date{Received July 2013 / Accepted 16.02.2014}

\abstract {Several recent studies have found that planet migration in
  adiabatic disks differs significantly from migration in isothermal disks. 
  Depending on the thermodynamic conditions, that is, the effectiveness of
  radiative cooling, and on the radial
  surface density profile, planets migrate inward or outward. Clearly, this
  will influence the semimajor axis-to-mass distribution of planets predicted by population-synthesis simulations.}
{Our goal is to study the global effects of radiative cooling, viscous torque desaturation, gap opening and 
  stellar irradiation on the tidal migration of a synthetic planet population.} 
{We combined results from several analytical studies and 3D hydrodynamic simulations in a new 
  semi-analytical migration model for the application in our planet population synthesis calculations.} 
 {We find a good agreement of our model with torques obtained in 3D
   radiative hydrodynamic simulations. A typical disk has three convergence
   zones to which migrating planets move from the in- and outside. This
   strongly affects the migration behavior of low-mass
   planets. Interestingly, this leads to a slow type II like migration behavior
   for low-mass planets captured in these zones even without an ad hoc
   migration rate reduction factor or a yet-to-be-defined halting mechanism.
 This means that the new prescription of migration that includes nonisothermal effects
  makes the previously widely used artificial migration rate reduction factor obsolete.}
{Outward migration in parts of a disk helps some planets to survive long enough to become 
  massive. The convergence zones lead to potentially observable accumulations
  of low-mass planets at certain semimajor axes. Our results indicate that
  more studies of the mass at which the corotation torque saturates are
  needed since its value has a main impact on the properties of planet populations. }

\keywords{Stars: planetary systems -- Stars: planetary systems: formation -- Stars: planetary systems: proto-planetary disks  -- Planets and satellites: formation -- Planets and satellites: migration -- Solar system: formation}  
 
\titlerunning{} 
 
\authorrunning{K.-M. Dittkrist et al.} 
                                    
\maketitle 
 
\section{Introduction}\label{sect:introduction} 
The huge diversity found in the properties of extrasolar planets is challenging
to reproduce with global theoretical planet formation models. The goal of such a model is to 
explain all the different planet types, which range from  low-mass rocky
planets such as Kepler-10 b 
\citep{batalha2011} and multiplanet systems like our solar system to high-mass planets orbiting far from their 
star, such as NR 8977 \citep{Maroisetal2008}. 
 
The only way to study this problem is to use the results of global formation and evolution 
models and to   
compare them statistically with the steadily increasing number of known planets and their physical properties. This is done in planet populations synthesis calculations, in which the evolution of one or several planets and the harboring protoplanetary disk is calculated at the same time in Monte Carlo simulations to create whole populations of planets. 
Several groups presented studies based on this method, for example, \citet{idalin2004},\citet{idalin2010}, 
\citet{Thommesetal2008},\citet{Miguelbrunini2008},\citet{HellaryNelson2012}, or by our group, 
\citet{mordasinialibert2009a},\citet{mordasinialibert2009b},\citet{mordasini2012a},\citet{mordasini2012b}, and \citet{alibertmordasini2011}.  
 
One general result found in all these models is that giant planets close to
the star (``hot'' Jupiters) do not form \textit{insitu}: the extrapolation of disk properties found at larger 
distances to small distances indicates that there is probably not enough solid material close-in to 
form a sufficiently large core that whould be able to accrete gas. The amount of material a core can 
accrete locally is given by the isolation mass $\miso$. According to the empirical 
minimum-mass solar nebula model (MMSN), the isolation mass is only a fraction of the 
Earth mass ($\mearth$) inside of 1 AU \citep{idalin2004}. Therefore an increase of 
solid matter by roughly two orders of magnitude compared with the MMSN would be 
needed \citep{idalin2004}. This means that to explain the close-in ``hot''
Jupiters, they would have to  
form initially at larger separations from the star and 
move inward by some mechanism (e.g., planet-planet scattering \citep{RasioFord1996}, Kozai mechanism 
\citep{NagasawaIda2008}, or tidal interactions with the gas disk 
\citep{goldreichetremaine1980,tanakatakeuchi2002}).  
In the planet formation model used in this work, we consider only one 
core per disk, which means that only migration caused by tidal interactions can be studied. This 
migration is generally described by two different regimes that depend on the 
mass of the protoplanet. The first is type I migration for low-mass planets, which are too 
small to form a gap in the disk, and the second is type II migration for
planets that open a gap \citep{DangeloHenning2002}.  
 
In our previous work
\citep[e.g. ,][]{mordasinialibert2009a,alibertmordasini2011}, we used the
results obtained for isothermal disks reported in 
\citet{tanakatakeuchi2002} to calculate 
type I migration rates. The migration rate in this model only depends on the disk 
surface density profile and not on the temperature profile of the disk because it 
assumes a globally isothermal disk. Migration rates obtained with this model
always lead to rapid 
inward migration in disks with profiles similar to the MMSN. 
We showed \citep{mordasinialibert2009a} that to obtain a synthetic population 
compatible with observations, one needs 
to artificially reduce the isothermal type I migration rate by a large factor. \citet{idalin2004} used a similar type I migration 
prescription and found necessary reduction factors of $\lesssim0.1$ for the migration rate.   
  
Recent studies of type I migration in 2D or 3D hydrodynamical simulations 
also found outward migration for some masses or semimajor axes, depending on the disk temperature structure 
\citep{massetdangelo2006,paardekoopermellema2008,kleybitsch2009}. More
analytical work derived a formulation that could be used in planet  
population synthesis calculations \citep{casolimasset2009,massetcasoli2009,massetcasoli2010}.  
Finally, \citet{paardekooperbaruteau2010} derived a formalism for type I 
migration in adiabatic or locally isothermal disks and improved this even more in \citet{paardekooper2011}.  
 
In the present work we describe a semi-analytical type I migration model that
can be applied to a wider range of planet and disk properties than that of
\citet{paardekooper2011}. For this we used the adiabatic and 
locally isothermal migration equations from \citet{paardekooperbaruteau2010} 
and ratios of relevant time scales to determine the transition between different 
regimes. We then studied the global consequences of the physics included 
in the migration model in new sets of population synthesis calculations. 
 
As an overview of this work, the new semi-analytical model we created is introduced in Section 2, where we 
also compare torques obtained with this model with torques 
obtained with the model of \citet{paardekooper2011} and data obtained in 3D radiative 
hydrodynamic simulations of \citet{kleybitsch2009} and \citet{bitschkley2011}. 
We discuss in Section 3 a reference synthetic planet population calculated with 
the nominal model, and in Section 4 we study the effect of different migration models  
on planetary populations.  
In Section 5 we draw our conclusions and summarize our results.

\section{Migration model} 
\label{migmodel} 
The migration module in our planet formation model \citep{alibertmordasini2004} 
distinguishes three main regimes, type I, 
disk-dominated type II, and planet-dominated type II migration \citep{armitagerice2005}. Low-mass planets up 
to typically a few 10 Earth masses migrate in the type I regime, followed by migration in 
disk-dominated type II regime for more massive planets, which finally pass into 
planet-dominated type II migration when they reach masses of typically 100-200 $\mearth$.  
We first describe our improvements to the description of the type II 
regime and introduce the new type I migration model afterwards. 
 
\subsection{Type II model  and outward migration} 
In the old model, the type II migration rate was calculated using the equilibrium 
flux of gas in the disk, which was always assumed to be directed inward 
\citep{mordasinialibert2009a}. Now the direction and  
rate of type II migration is numerically calculated by considering the radial  
velocity $v_{\rm gas}$ of the (nonequilibrium) flux of gas at the position of the 
planet. It therefore allows outward migration if the planet is 
in a part of the disk where the gas is flowing outward 
\citep{verasarmitage2004}.  The planetary migration rate is then 
\beq 
\dot{a}_{\rm p, T2}=v_{\rm gas} \times  \rm{Min}(1, 2 \Sigma \rp^{2}/\mplanet) ,
\eeq 
where $\rp$ is the semimajor axis of the planet, $\mplanet$ is its mass, and 
$\Sigma$ is the gas surface density.   
 
This mechanism has been invoked to explain the formation of exoplanets with
semimajor axes larger than 20 AU, which cannot have formed in situ via the core accretion model \citep{verasarmitage2004}. 
However, we find in the syntheses using the nonequilibrium model that outward migration in 
type II is seldom important, and no large-scale net outward migration over 
more  than $\sim 1\ AU$ typically occurs due to it. The reason is twofold:   
 
The radius of velocity reversal (or of maximum viscous couple) $R_{\rm MVC}$ 
\citep{lynden-bellpringle1974} is relatively close-in only early 
in the disk evolution. But, at these early times, the planets have usually not 
yet grown to a mass regime in which they migrate via type II. The 
evolution of the disk leads to the subsequent recession 
of $R_{\rm MVC}$ to larger radii. This occurs faster than the growth of the 
planets. Therefore type II outward migration 
is a very rare event during the spreading phase of the disk: at the moment planets 
have grown massive enough to migrate in type II, they are most of the time already  
located inside  the $R_{\rm MVC}$.  
 
Another chance of outward migration exists towards the end 
of the disk lifetime, when parts of the disk flow outward because mass is removed 
at the outer border due to external photoevaporation.  The gas surface densities, however,
are typically already quite low at the position of the planet at 
this moment, so that the reduction factor of the planet's migration rate 
relative to the viscous velocity $\propto \Sigma a_{\rm planet}^{2}/M_{\rm 
  planet}$ \citep[see][]{alibertmordasini2005} is low as well, 
leading again to only modest amounts of outward migration. As in the original 
model, we assumed a linear reduction factor if $M_{\rm 
  planet}>2\Sigma a_{\rm planet}^{2}$ (``fully suppressed'' planet-dominated
type II), because this agrees a better with hydrodynamical
simulations than a square-root dependence on the planetary mass
\citep{alexanderarmitage2009}.  Outward  
migration during effective photoevaporation is additionally limited because
the remaining disk-lifetime is short.   
 
\subsection{Type I migration regimes} 
\label{theory} 
Here we discuss the migration of low-mass planets below a few tens of Earth masses. 
As mentioned above, one of the problems with the original description of orbital migration of low-mass 
planets is the short time scale found in isothermal type I migration
\citep{tanakatakeuchi2002}, which resulted in too many close-in planets in
planet population synthesis calculations (see also the comparison in Section
\ref{oldmodel}). These rates had to be artificially reduced by correction
factors to produce enough ``cold'' giant planets at larger semimajor axes
\citep{mordasinialibert2009b,idalin2008} to fit the observations.   
 
On the other hand, \citet{PaardekooperMellema2006} and \citet{kleybitsch2009} 
showed that migration of small-mass bodies is not always inward because in a nonisothermal disk 
migration can be directed outward for some 
masses. Outward migration in the Type I regime can also occur in isothermal
disks if full MHD turbulence is considered \citep[see
also][]{UribeKlahr2011,GuiletBaruteau2013}, but these new effects are not 
considered here because they need to be studied in more detail more studies before they can be parametrized.  
 
Recently, \citet{paardekooperbaruteau2010} derived semi-analytical 
formulas for migration in the limiting case of adiabatic disks. Here we
combine different formulas that are valid in different 
thermodynamical regimes into a model that can be applied to the wide 
range of planet masses and thermodynamic properties of the disk that are needed for our population-synthesis models. 
 
\subsubsection{Type I fit formula} 
 
\begin{figure} 
         \centering \includegraphics[width=0.75\linewidth]{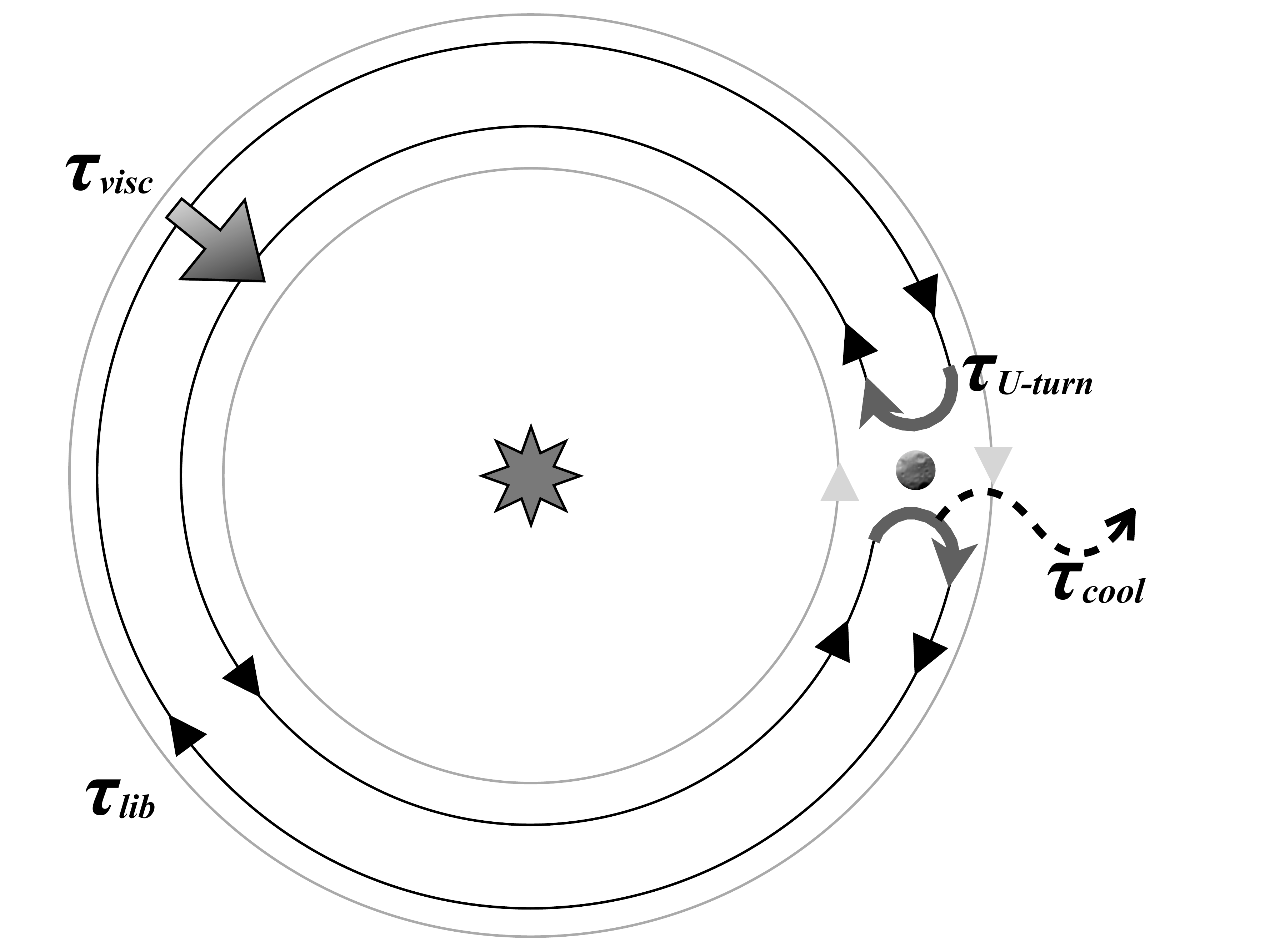} 
	\caption{Schematic representation of the relevant time scales involved 
          in type I migration. Flow lines of gas parcels are indicated in 
          a system of reference that rotates with the planet. The planet is indicated on 
          the right. In the center is the star. Close to the planet, the 
          flow lines bend and make a u-turn during a time equal 
          $\tauuturn$. During that time, the gas on the flow lines cool on a time scale 
          $\taucool$. One full libration around the planet (indicated by the 
          black lines) takes a libration time, $\taulib$. During this time, 
          viscosity acts on a viscous time scale $\tauvisc$. The corotation 
          region lies between the two black lines. Inside and outside this 
          region, gas parcels do not make u-turns, but have a velocity 
          relative to the planet because of to the Keplerian sheer (gray lines).}  
	\label{schematictimescales} 
\end{figure}

Nonisothermal migration rates are more complex than isothermal rates.  
Generally, the gravitational interactions of the planet and the gas disk can lead to 
three characteristic flow regions that produce different types of torques:  
 
\begin{itemize} 
\item{Lindblad torques:} 
 
Gas sufficiently far from the planet is only slightly 
perturbed and orbits the star on nearly circular orbits inside or outside the planet's 
orbit. The gravitational interaction generated by the planet in the regions outside 
and inside the corotation region produces the Lindblad torques.  
 
\item{Corotation torque:} 
 
If the orbit of a gas parcel is closer to the planet, its flow is more and 
more deflected, until it passes the planet's orbit in front or behind the 
planet. Thus it forms so-called horseshoe orbits (see Fig. \ref{schematictimescales}). 
The gas in the horseshoe orbits produces by deflection the 
so-called horseshoe drag or corotation torque. The strength of this torque depends on the 
thermodynamic regime of the interaction between the planet and the disk and 
the mass of the planet.  
\end{itemize}

For typical properties (radially decreasing density and temperature) of the disk the Lindblad torques lead 
to inward migration, whereas the isolated torques from the corotation region 
can result in either inward or outward migration. For certain thermal and 
surface density profiles in the disk, these torques can be 
stronger than the Lindblad torques. Thus the combination of Lindblad and 
corotation torques can lead to either inward or outward migration, depending on 
their relative strength, which is determined by the disk properties.  
 
The total torque $\Gamma$  can be expressed in the following way, as shown in \citet{tanakatakeuchi2002}, \citet{paardekooperbaruteau2010} or \citet{massetcasoli2010}:
\begin{eqnarray}  
\tilde\Gamma = \frac{\Gamma}{\mplanet} & = & \tilde{C} \rp^2 \omegap^2  \rm{ \
  \ \ , where} \\ 
\tilde{C} & = & C \frac{\rp^2}{h^2} \sigmap \frac{\mplanet}{\mstar^2} = 
\frac{\dot{a}_p}{2 \rp \omegap} = \frac{1}{4 \pi N_{\rm orb}} =
\frac{\tau_{\rm{orb}}}{4\pi \tau_{\rm{mig}}} , \label{eqa} 
\end{eqnarray} 
In these equation, $\tilde\Gamma$ is the specific torque (torque per unit 
mass). With a small p we denote all quantities of or at the position of the 
planet. $\mplanet$ is the total mass of the planet and $\rp$ the
semimajor axis, while $\mstar$ is the mass of the star, $h=H/a$ is the aspect ratio of 
the disk with a vertical scale height $H$. $\omegap$ is the Keplerian 
frequency of the planet and $\sigmap$ the gas surface density.  The dimensionless 
factor $C$ in the second part of Eq. \ref{eqa} gives the direction and strength of the migration and is discussed 
below for different thermodynamical regimes (Sect. \ref{theory}). The dimensionless number $\tilde{C}$ is proportional to the 
migration rate $\dot{a}_p$ and inversely proportional to the number of orbits needed for a 
planet to migrate over the distance of its semimajor axis $N_{\rm orb}$. It is also proportional to the ratio of the two time scales for migration $\tau_{\rm{mig}}$ and orbital motion $\tau_{\rm{orb}}$ 
 
Additional parameters in the following equations are $\betas$ as the local power-law exponent of the gas surface density ($\Sigma \propto r^{\betas}$), $\betat$ the local power-law exponent 
of the temperature ($T \propto r^{\betat}$), and $\gamma$ the adiabatic index 
(ratio of the heat capacities) of the gas. 
 
Depending on the properties of the disk, $C$ is a combination of various torque contributions of variable importance. Before we describe how we combined the contributions, we introduce expressions for their individual strength.

\citet{paardekooperbaruteau2010} derived that the Lindblad torque in an 
adiabatic disk is proportional to (their Eq. 47, part 1) 
\beq 
\clind = \frac{1}{\gamma}\left(-2.5 +1.7\betat - 0.1\betas\right). 
\label{clind} 
\eeq 
They also found that the horseshoe drag in the adiabatic case is proportional 
to (their Eq. 47, part 2) 
\beq 
\chsa = \frac{1}{\gamma}\left(1.65 +\betas (9-7.9/\gamma)- 7.9\betat/\gamma \right). 
\eeq 
The coefficient $C$ in the adiabatic regime  due to the combination of the Lindblad and 
corotation torques is 
\beq 
\cadia = \clind + \chsa \label{chap2 eq:cadia}. 
\eeq 
\citet{paardekooperbaruteau2010} also found that the total torque in a 
locally isothermal regime, where the temperature $T$ is constant in time but not 
with semimajor axis, is proportional to (their Eq. 49) 
\beq 
\cloc=-0.85+0.9\betat+\betas \label{chap2 eq:cloc}. 
\eeq 
Subtracting from this the Lindblad torque in the adiabatic regime, but setting 
$\gamma = 1$ (compare \citet{paardekooperbaruteau2010} Sect. 5.4 \footnote{One infers the locally 
  isothermal regime by taking the limit $\gamma \rightarrow 1$, which invokes 
  infinitely efficient thermal diffusion.}), one finds the horseshoe drag part in 
the locally isothermal regime as 
\beq 
\chsl=1.65-0.8\betat+1.1\betas \label{chap2 eq:chsl}. 
\eeq 
Compared with  Eq.\ \ref{clind}, \citet{massetcasoli2010} derived a partially different Lindblad torque. 
We study the effect of this weaker Lindblad torque in Sect. \ref{sec:Lindblad Torques}. 
\beq 
C_{\rm Lind,2} = \frac{1}{\gamma}\left(-2.5 +0.5\betat - 0.1\betas\right). 
\eeq

\subsection{Time Scales} 
\label{timescales} 
The proper mix of the above described torque contributions can be determined by investigating the relevant time scales. For instance a disk behaving adiabatically produces a different torque than a locally 
isothermal one. Here the relevant time scales are the cooling time in comparison to the dynamic time. 
 
To decide in which subtype the planet belongs to, 
we compared four characteristic time scales in total. The different time scales are 
schematically shown in Figure \ref{schematictimescales}. In all the following 
estimates of the time scale, the important characteristic length scale is the width of the horseshoe 
region  $\xs$ given as 
\citep{massetdangelo2006,baruteaumasset2008,paardekooperbaruteau2010}  
\beq 
\xs = 1.16 \rp \sqrt{\frac{q}{h\sqrt{\gamma}}}. 
\eeq 
In this equation $q$ is the ratio of the planet mass to the stellar mass. 
 
The first two time scales we compared are the cooling time and the u-turn time to distinguish between the locally isothermal and the adiabatic regime. The u-turn time is 
the time a gas particle needs to undergo one turn in front or behind the 
planet. Its value is approximately given as \citep{baruteaumasset2008}  
\beq 
\tauuturn  =   \frac{64 \xs h^2}{9 q \rp \omegap}. 
\eeq 
The cooling time of a gas blob undergoing a turn is calculated by solving the 1D equation \citep{kleybitsch2009} 
\beq 
\frac{dT}{dt}=-\frac{1}{\rho C_{\rm V}} \frac{\partial}{\partial a}\left(F\right), 
\eeq 
where $F$ is the heat flux, $\rho$ the gas density, and $C_{\rm V}$ the heat 
capacity at constant volume. We assumed a cooling over the length $\lcool$, 
which is the minimum of $H$, and $\xs$, which corresponds to either horizontal cooling over 
the width of the horseshoe region or vertical cooling through the disk.  
The flux $F$ in the optically thin case, when $\rho \kappa \lcool <
\sqrt{1/8}$ is $F = \tau \sigma T^4$, and in the optically thick case in the
diffusion description $F= \frac{4 a c T^3}{3 \rho \kappa} \frac{\partial
  T}{\partial a}$ \citep{kleybitsch2009}.
This means that we have two different types of the cooling time scale with a smooth
transition when the optical depth is $1/\sqrt{8}$  
\beq 
\taucool = \frac{\lcool \rho C_V}{8 \sigma T^3} \left(8 \rho \kappa \lcool + \frac{1}{\rho \kappa \lcool}\right). 
\eeq 
 
In a similar way we compared the viscous time scale and the libration time scale 
to find out whether the horseshoe drag is saturated or not \citep{massetcasoli2010}. 
The viscous time scale is  
\beq 
\tauvisc = \frac{\xs^2}{\nu}, 
\eeq 
while the libration time is given as \citep{baruteaumasset2008}
\beq 
\taulib  = \frac{8 \pi \rp}{3 \omegap \xs}. 
\eeq 
 
We denote the ratios of the relevant time scales as 
\beq 
s_1=\fcool \frac{\taucool}{\tauuturn} 
\eeq 
and  
\beq 
s_2=\fvisc \frac{\tauvisc}{\taulib}. 
\label{s2fvisc} 
\eeq 
An equivalent approach can also be found in \citet{casolimasset2009} and \citet{massetcasoli2009}. 
 
These time scales are typically order-of-magnitude estimates. Therefore, we 
introduced two factors, $\fcool$ and $\fvisc$, of order unity to adjust the point of 
transition between different regimes to obtain a better agreement with radiative hydrodynamic 
simulations. In this work we set $\fcool= 1.0$, since this proved toagree well
with numerical results of migration rates, as we show in  
Section \ref{compmodel}. There, we also study the influence of various values of 
$\fvisc$. In general, we find that in nonirradiated $\alpha$ 
disks, planets transit from the locally isothermal into the adiabatic regime before the 
corotation torque saturates (Sect. \ref{refsyncalc}). 

We also note that for a given $h$, the ratio of the u-turn time and libration time is constant:
\beq
\tauuturn=\frac{1.16^2 8}{3 \pi}\frac{h}{\sqrt{\gamma}}\taulib \approx 1.14
\frac{h}{\sqrt{\gamma}}\taulib .
\label{uturnlibratio}
\eeq
  
\subsection{Type I migration formula} 
To combine the migration rates of these different regimes we defined 
an arbitrary transition function $z$  to obtain a smooth shift from the locally isothermal to the adiabatic regime as a function of 
the variable $s_1$: 
\beq 
z(s_1)= \frac{1}{1+s_1^b}, \label{transfunc}  
\eeq 
which has the properties that for $s_1 \rightarrow 0, z(s_1)\rightarrow 1$ and for 
$s_1 \rightarrow \infty, z(s_1) \rightarrow 0$. Furthermore, depending on the value of 
$b$, the transition from 1 to 0 occurs more or less quickly around $s_1=1$. 
Therefore, uncertainties in overlap of different regimes can be approximated with a lower $b$ 
value. Additionally, the continuity of the transition function allows one to use 
longer numerical time steps in a simulation when the planet is close to the transition from one regime into another.  
 
For individual transitions (e.g., locally isothermal to adiabatic, or type I 
into type II), other studies (\citet{paardekooper2011}; \citet{massetcasoli2010}) derived  
physically motivated transition functions. But the comparison of physically motivated 
transition functions \citet{paardekooper2011} with our simple transition function 
defined above shows little difference in population synthesis models (see 
Section \ref{paardepopsyn}). We set $b=4.0$, but 
the actual value of $b$ is again not very important for the global outcome seen 
in a population if $1.5\lesssim b \lesssim 100$ (see Appendix \ref{minoreffects}). 
   
We multiplied the horseshoe part with $\min(1/s_2,1)$ to account for the
saturation of the torque that originates
in the horseshoe region. As shown in previous studies (\citet{masset2002} and \cite{massetcasoli2010}), even when 
the cooling time is shorter than the u-turn time, and thus it is also much shorter than the libration time, viscosity 
is the driving factor for the saturation of the horseshoe region.  This yields the following 
type I migration coefficient $C_{\rm T1}$ (in the second term of Equation \ref{eqa}):

\beq 
C_{\rm T1} = \clind + \min\left(1/s_2,1\right) \left( z(s_1) \chsl + \left(1-z(s_1)\right)  \chsa \right).
\label{ct1v2} 
\eeq

This assumes that even when cooling is much faster than the libration time
scale  the horseshoe drag will saturate without sufficient 
viscous coupling of the horseshoe region to the rest of the disk. The horseshoe part itself shifts between the adiabatic and the
locally isothermal value depending on the ratio $s_1$ of the cooling time scale 
to the u-turn time scale (if $h$ is constant also with respect to the
libration time scale (Eq. \ref{uturnlibratio})). 
 
The reduction of the surface density at the planet's location due to the beginning gap 
formation for more massive planets will also modify the migration behavior. \citet{cridamorbidelli2007}  
derived a formula to estimate the depth of the gap relative to the unperturbed 
disk. In their definition, a gap is formed when the surface density is reduced to 10\% of the 
unperturbed value. They defined a factor (their Eq. 12) 
\beq 
P_{\rm \Sigma} = \frac{3 h}{4\sqrt[3]{q/3}} + \frac{50 \nu}{q \aplanet^2 \omegap}  
\eeq   
as a combination of two criteria (the thermal and the viscous condition) and used this factor 
to approximate the depth of the gap as 
\beq 
\fsigma(P_{\rm \Sigma}) =  
\begin{cases} 
\frac{P_{\rm \Sigma}-0.541}{4.0} & \text{if $P_{\rm \Sigma} < 2.4646$ ,} \\ 
1.0-\exp(\frac{P_{\rm \Sigma}^{3/4}}{-3.0}) & \text{otherwise .} 
\end{cases}  
\eeq 
We multiplied $\sigmap$ in Eq. \ref{eqa} with this factor to reduce the surface 
density in our torque calculations. This means that  the type I migration rate 
is reduced by a factor of up to 10 at the point of transition to type II 
migration. This reduced type I migration rate is still about one order of 
magnitude higher than the type II migration rate for typical disk conditions 
(Sect. \ref{compmodel}), which necessitates using a smooth transition function necessary. 
 
Eventually, when the planet mass reaches the gap opening mass $\mgap$, that is, the 
mass at which $\fsigma = 0.1$, the torque transitions to type II: 
\beq 
C=z\left(\frac{\mplanet}{\mgap}\right) C_{\rm T1} + 
\left(1-z\left(\frac{\mplanet}{\mgap}\right)\right) C_{\rm T2} .
\label{cfinal} 
\eeq 
We used the same functional dependency as in (Eq. \ref{transfunc}) for a smooth transition 
from the type I to the type II torque. Here, we set the smoothing parameter $b=10.0$ (fast transition), because the reduction of the surface density already partly smoothes the  
transition (see Sect. \ref{minoreffects}). The transition factor $s_{\rm type II}$ is in this case the ratio of the planet mass to the mass at which the planet would open a gap in 
the disk:
\beq 
s_{\rm type II}=\frac{\mplanet}{\mgap}. 
\eeq 
The factor $C_{\rm T2}$ is the corresponding type II migration coefficient 
in the same dimensions as $C_{\rm T1}$, given as  
\beq 
C_{\rm T2}=\frac{\dot{a}_{\rm p, T2} \mstar h^2}{2 \rp^3 \sigmap q \omegap}. 
\eeq 
 
\subsection{Comparison with other models}  
\label{compmodel} 
 
In this section we compare our model with that of \citet{paardekooper2011} for
a specific choice of parameters. The global consequences in a full planetary
population are shown in Section \ref{paardepopsyn}.  Here, we use a simple, nonevolving 
1D-disk and compare the torques predicted by the two models in such a disk with results 
from the 3D radiation-hydrodynamic simulations of \citet{kleybitsch2009} and 
\citet{bitschkley2011}. 
 
\citet{paardekooper2011} developed a more complex migration model trying to use first principles to derive in particular the transition functions compared with our model 
described in Sect. \ref{migmodel}. They derived their torques by calculating
the effect of viscosity and thermal diffusion onto the horseshoe region itself. From this they found 
transition functions between different parts of the corotation torque, using thermal 
diffusion time scales and viscous time scales as transition criteria between 
barotropic and entropy-related parts of the horseshoe drag and the linear 
corotation torque and also the saturation of both. On the other hand, we altered our torque
according to the ratio of these time scales using an arbitrary transition. 
They called their ratio of these time scales $p_\nu$ and $p_\chi$ with $p_\nu^2$, the ratio scaling the saturation due to viscosity, being equal to our $s_2$ ratio if $\fvisc=0.75$.
For $p_\chi$ one can show that
\beq
\frac{\pi}{3} p_\chi^2 = \frac{1.16^2 8}{9} \frac{h}{\sqrt{\gamma}} s_1.
\eeq
The numerical factors in front of both sides of the equation are
  almost equal 1. The transition functions in \citet{paardekooper2011} and our
  work are different. Therefore we considered the turnover points to compare the transitional behavior. In our model it
  occurs when $s_1=1$. For typical values of $h$ and $\gamma$ this leads to a
  value of $p_\chi \approx 0.2$. While not as easy to characterize
  due to the multiplication of several factors (see their Eqs. 23, 30, 31, and
  53), this is in the range where the transition in the entropy torques occurs also in the
  model of \citet{paardekooper2011}.
The main difference in the description is the separate treatment of the "entropy" and "barotropic" 
parts in \citet{paardekooper2011} with three slightly different transition
  functions between different parts of the torque. We only used one transition
  function between the locally isothermal and adiabatic regime to attribute
  the dependence of the torque on cooling processes. Finally, they used only one free parameter, the smoothing factor for the 
planetary potential, which we set to 0.4 in our comparison here.   
They also stated that their model is mostly applicable to planets with a mass
of a few Earth masses, while our model is intended to be applicable to a wider mass
range.

To quantify the impact of the differences, we next compared the torques
predicted by our simple model with those in 3D radiation hydrodynamic
simulations, and also  with the results found with the Paardekooper model. As
will become clear, one finds that despite the simplifications in our model, we
can fit the numerical results relatively well.

The 1D disk for the comparison was set up to be as similar as possible to the 
disk from the 3D simulations of \citet{kleybitsch2009} and 
\citet{bitschkley2011}. The surface density $\Sigma$ is a power-law in semimajor axis 
with a fixed slope of -0.5, and the temperature $T$ is a power-law in semimajor axis of 
-1.5 inward of 2.5 $a_{\rm Jup}$, where $a_{\rm Jup}$ is the semimajor axis of
Jupiter. Farther out, when the temperature almost reaches 10 K, its power-law exponent $\betat$ increases and approaches
0. We used the same EOS \citep{SaumonChabrier1995} and opacity tables \citep{belllin1994} as in our full model. The viscosity was set to the same value as in
\citet{kleybitsch2009}. All other disk parameters, such as disk height $H$ or 
density $\rho$, were calculated from $\Sigma$ and $T$. In this disk we then calculated the specific torques on 
planets of either different mass or different semimajor axes. 
With $\betas = -0.5$ and 
$\betat=-1.5$ there will be no type II torque since there is no radial gas 
flow in this disk. We stress that our disk model for the planet population synthesis  
calculations is in contrast a time-evolving 1+1D $\alpha$ model and is summarized in Sect. \ref{sect:diskmodel}. 
 
\begin{figure} 
         \centering \includegraphics[width=1.0\linewidth]{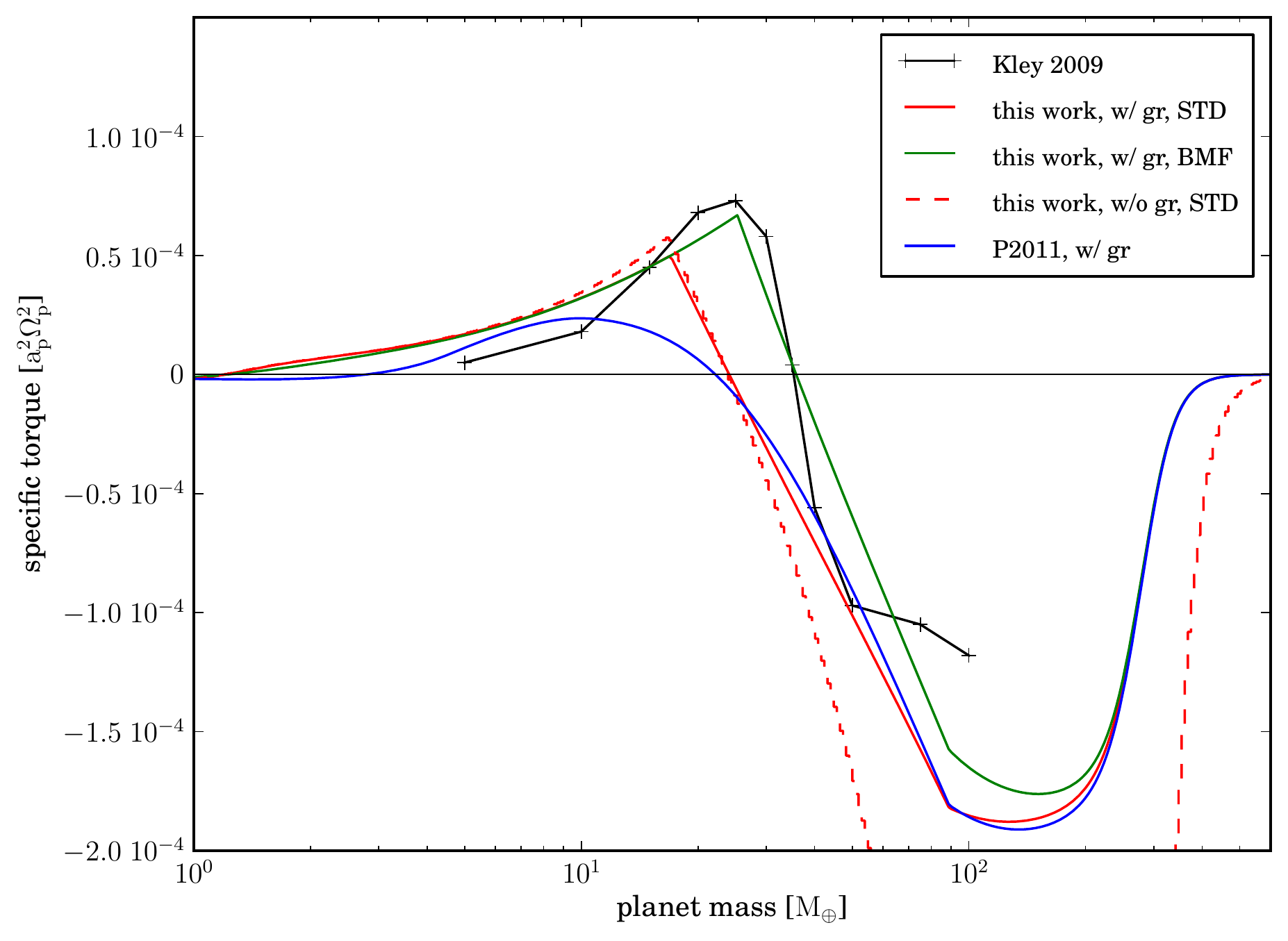} 
	\caption{Specific torque for different planet masses in the range 
        from 1 to 600 $\mearth$  at 5.2 AU. The red solid line shows torques 
          obtained with our nominal case ($\fvisc = 1.0$), the green line is 
        from the BMF model ($\fvisc = 0.55$). The red dashed line does not include the 
        reduction of the surface density due to gap formation (gr), otherwise it is 
      identical to the nominal model. The black line with crosses shows
      torques from the  
    3D-radiative hydrodynamic simulations of \citet{kleybitsch2009}. A blue line  
    shows torques obtained with the model of \cite{paardekooper2011}. 
	The solid green line is our model of choice for the population synthesis 
     models because it fits the results of full 3D simulations best.} 
	\label{fig1} 
\end{figure}

\begin{figure} 
         \centering \includegraphics[width=1.0\linewidth]{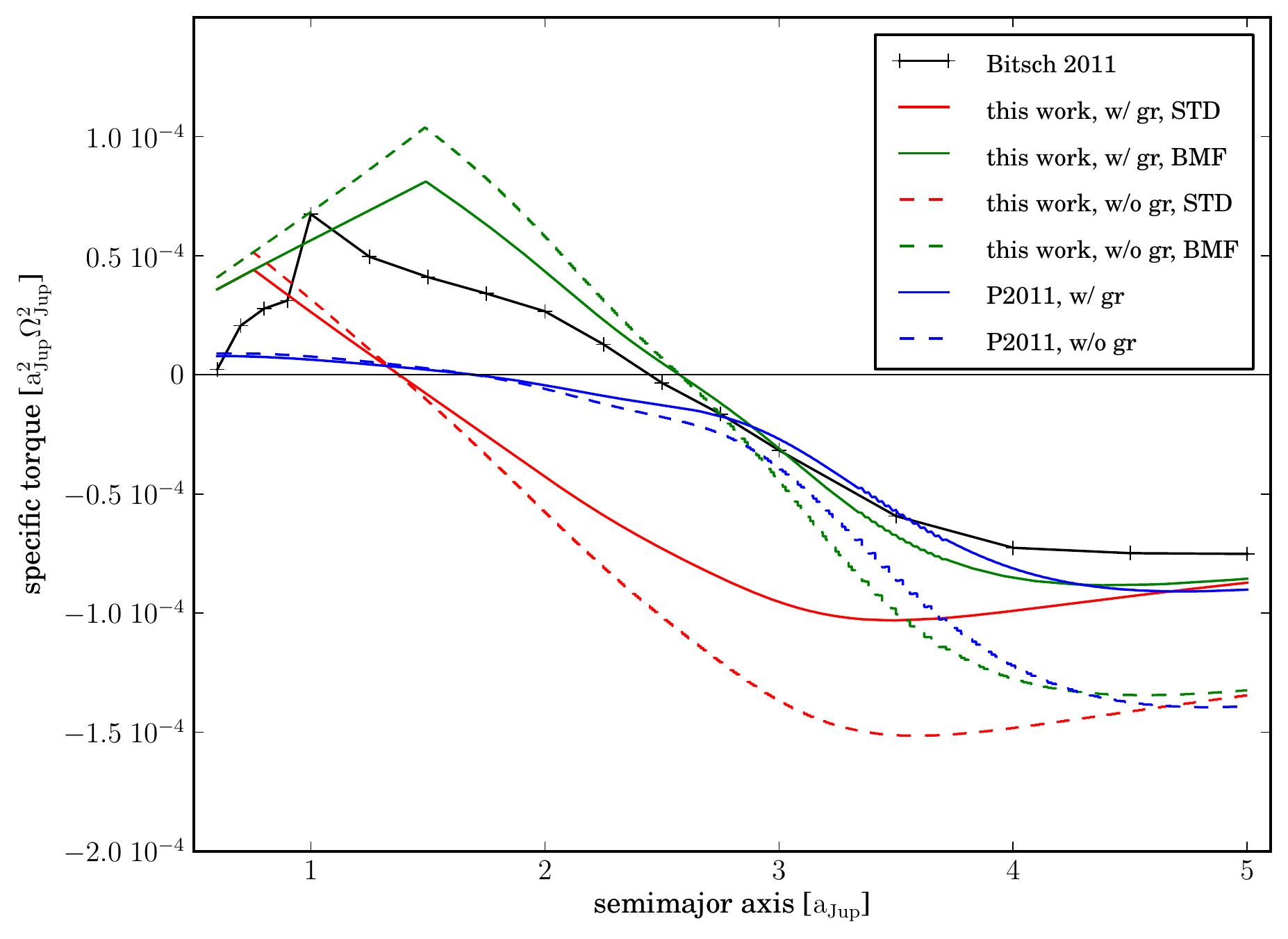} 
	\caption{Specific torque for different semimajor axes in units of the 
          semimajor axis of Jupiter $a_{\rm Jup}$ for a 20 Earth 
          mass planet. The red lines show torques 
          obtained with our STD model ($\fvisc = 1.0$), the green lines 
          with the BMF model ($\fvisc = 0.55$). The solid (dashed) lines do (not) include the 
        reduction of the surface density due to gap formation (gr). The line
        with black crosses shows torques found in the 
    3D-radiative hydrodynamic simulations of \citet{bitschkley2011}. The blue line  
    shows torques obtained with the model of \cite{paardekooper2011}. The solid green line is our model of choice for the population synthesis 
     models because it fits the results of full 3D simulations best.} 
	\label{fig2} 
\end{figure}

In Fig. \ref{fig1} we plot torques from our model as a function of planetary mass 
and compare them with simulations from \citet{kleybitsch2009}, while in Figure \ref{fig2}, we compare our model 
with \citet{bitschkley2011}, where torques at different semimajor axes for a
20 $\mearth$ planet were calculated. In both figures we also show  the torque as predicted 
from the model of \citet{paardekooper2011} (for their suggested optimal free parameter $b=0.4$). 

We plot the migration rates for our model with and without 
the reduction of the surface density due to gap formation (solid and dashed
lines respectively). This reduction was not only applied to our own model but
also to the \citet{paardekooper2011} prediction. For the dependency of planet masses
(Fig. \ref{fig1}), the curves with the gap 
reduction remain closer to the data obtained by \citet{kleybitsch2009}. At 
around 200 $\mearth$ when the transition to type II occurs, this factor 
causes a 10 times lower type I torque than without the reduction. Therefore 
with the zero type II torque in this disk-setup, the overall torque is also 10 
times smaller.  
The red lines correspond to our model described above (with $\fvisc = 1$: hereafter the STD "standard"-model), while 
the green lines are our model with $\fvisc = 0.55$ (hereafter the 
BMF "best mass fit"-model). The latter value for  $\fvisc$ was chosen to increase the saturation
mass in a way so that the mass of zero torque in our own semi-analytical model
agrees  with the data of \citet{kleybitsch2009}. The associated reduction of
transition parameter $s_2$ could be interpreted as a more efficient viscous injection of angular
momentum into the horseshoe region from the rest of the disk than in the
STD-model. Only a factor $\sim2$ is needed to bring our model in agreement
with the hydrodynamic simulations, meaning that the simple estimate of the transition using the
characteristic time scales leads to relatively accurate results.    
Setting $\fvisc = 0.55$ also increases the maximum specific torque, which
follows the data from \citet{kleybitsch2009} more closely than our model
with  $\fvisc = 1$, or the model of  \citet{paardekooper2011}, at masses
larger  than 15 $\mearth$.   

In Figure \ref{fig2} we compare torques on a planet of a fixed mass (20
$\mearth$) as a function of semimajor axis. Here we see as well that the
BMF-model provides the best agreement with the data of \citet{bitschkley2011}.
The semimajor axis at which the torque vanishes is with 2.7 $a_{\rm 
  Jup}$ relatively close to the 2.5 $a_{\rm Jup}$ found in the hydrodynamical 
simulations. The model of \citet{paardekooper2011} places the position of zero torque at 1.8 $a_{\rm 
   Jup}$, while in the STD-model it occurs at 1.4 $a_{\rm Jup}$.   
Thus, for planets  closer than 2.5 $a_{\rm Jup}$ the BMF-model seems to be
the best representation of both sets of numerical simulations. 

At larger distance from the star the BMF-model and \citet{paardekooper2011}
again yield similar results and compare relatively well to 
the data. All curves also indicate that including the reduction of the 
surface density in Eq. \ref{eqa} due to gap formation leads to a better 
agreement with the torques found in the hydrodynamic simulations, especially 
in the outer parts of the disk. In summary, we see that by setting $\fvisc = 0.55 $, 
the model agrees fairly well with the data provided from 3D simulations.

Some diskrepancy exists for the innermost point considered in the hydrodynamic
simulation. It is most likely caused by effects of the change of the opacity due to
ice evaporation, which is handled in a different way in our model vs. the 3D full models, so
that the gradient of temperature and surface density differ.    

Note that this comparison in principle only applies for a fixed value of the
Prandtl number. The latter depends on $\gamma$ and the optical depth $\tau$
(\citet{paardekooper2011}), which varies, for instance with the distance from the star or temperature. In the limit of high optical depth, the Prandtl number approaches $Pr= \frac{9}{4}\gamma(\gamma-1)$.

Nevertheless, we conclude that our BMF-model can quite adequately reproduce the current understanding of planet migration based on (semi-) analytical models for the modeling purpose of planetary populations.
  
\subsection{Model versions}
\label{modelversions} 
In the last section, we have calibrated our semi-analytical model with the results of 
a specific set of radiation-hydrodynamic simulations. With these results, we define three versions
of the model that are used below to study the (global) effects of these different variants of our migration model: 
\begin{itemize} 
\item \textbf{STD model} In the standard version we simply set $\fvisc=1.0$ so that 
  the plain time scales as defined in Section \ref{timescales} are 
  employed to calculate the point of saturation. 
\item \textbf{BMF model} In the best mass fit version we multiply $s_2$ by
  0.55 ($\fvisc = 0.55$). This model compares best with the 3D migration simulations as shown 
above.  
\item \textbf{RED model} In the reduced version we reduce $s_2$ by multiplying 
it with 0.125 ($\fvisc = 0.125$). This results in a four times higher
saturation mass than for the STD case. We use this even larger reduction to study the effect of different reduction factors.  
\end{itemize}

\section{Formation models and migration tracks}  
In addition to the comparison shown above for the migration rates alone, we are
interested in the global effects of the new migration model onto a planet population compared with our 
older isothermal migration model. 
The planet formation model used for this is based on the paradigm of core accretion \citep{perricameron1974,mizunonakazawa1978,pollackhubickyj1996,alibertmordasini2005}.  
The model is described in detail in \citet{alibertmordasini2005} with later
modifications shown in \citet{mordasinialibert2009a}, \citet{mordasiniklahr2010,mordasini2012b,mordasini2012a},
and  \citet{fouchetalibert2012}.  
It consists of different computational modules, namely the planet accretion
module, the disk module, and the migration module. 
The migration module is described above and was used 
in recently published work of \citet{Fortier2013}, 
\citet{mordasini2012b}, and \citet{mordasini2012a}. 
In the following sections we give a short overview of the disk and 
accretion modules. Then we present calculations of a few specific planets, show the 
associated formation tracks and the important features found in them. Finally,
we present results from planet population synthesis calculations. 
Since we concentrate in this paper on the direct effects of the new migration
model, we present here only simulations with just one embryo per protoplanetary disk. The interplay of migration and multiple concurrently forming planets \citep{Alibert2013} will be addressed in future work.

\subsection{Disk model} 
\label{sect:diskmodel} 
As a model for the protoplanetary disk, we used a 1+1D model as in
\citet{papaloizouterquem1999} or \citet{belllin1994}, and present 
only a short summery here. The protoplanetary disk is described as a 
viscously evolving $\alpha$ disk, where $\alpha$ is assumed 
to be constant throughout the disk. The viscosity is given as $\nu=\alpha H \cs$ with 
$H$ the disk scale height and $\cs$ the sound speed \citep{shakurasunayev1973}. Irradiation effects 
from the host star can be included in the calculations of the vertical 
structure \citep{fouchetalibert2012}. For the evolution of the gas surface 
density $\Sigma$ over time $t$ and distance $a$ from the star we solved the 
standard viscous evolution equation from \citet{lynden-bellpringle1974}. For most of our simulations in this paper we neglected stellar irradiation and used only viscous heating if not otherwise mentioned:  
\begin{equation} 
\frac{d\Sigma}{dt} = \frac{3}{a} \frac{\partial}{\partial a}\left[a^{1/2}\frac{\partial}{\partial a}\left(\nu \Sigma a^{1/2}\right)\right] + \dot{\Sigma}_{\rm w}(a).
\end{equation} 
As a sink term we included the photoevaporation rate $\dot{\Sigma}_{\rm w}(a)$ as 
given in \citet{mordasinialibert2009a}. Together with $\alpha$ 
and the initial disk mass it determines the disk lifetime. At the start, the
solid surface density of the planetesimals is equal to the gas density times
the dust-to-gas ratio $\fpg$.  It is further reduced inside of the iceline at
a temperature $T=170\, \rm{K}$ by a factor of 4. Other than by accretion onto
and ejection by the planet, we did not evolve the solid surface density.

\subsection{Model of accretion and internal structure} 
\label{accretionandinternalstructure} 
We simulated the growth of one planet per disk. For this, we inserted a
$0.6\,\mearth$ seed embryo at a random position in the disk. The core has a
constant density of $3.2\,\rm{g/cm^2}$ and contains all  
the heavy matter the planet accretes, that is, we assumed that all planetesimals 
sink to the core \citep{pollackhubickyj1996}. The seed will initially start to accrete 
mostly planetesimals, which leads to a growth of the core. The amount of energy 
released from the infalling planetesimals is high at this point and the core 
mass is low, therefore initially only small amounts of gas are bound.   
As the core grows, it binds an increasingly massive  envelope. The accretion rate of gas is found by solving the standard equations of the structure of planet interiors, but with the simplification that the luminosity is constant throughout the envelope. 
 
In the original model of \citet{mordasinialibert2009a}, the luminosity of the planet is due
to the accretion of planetesimals alone. This is usually the dominant source of
luminosity at  smaller masses \citep{pollackhubickyj1996}. Here, we adopted a
simplified version of the model described in \citet{mordasini2012a} to take also
into account the luminosity generated by the accretion of gas. In the
original model it was sufficient to consider only the luminosity due to the
planetesimals, because the migration was always directed inward. This means that
the cores always migrate into regions with a high solid surface density. With
the new migration model, this is no longer the case: because the positive torques
act at certain masses (see Section \ref{compmodel}), it is possible that a
core migrates through parts of the disk with a very low solid surface density content. There the luminosity of gas accretion becomes important.

\subsection{Convergence zones} 
\label{sect:convzone} 
\begin{figure} 
        \centering \includegraphics[width=1.0\linewidth]{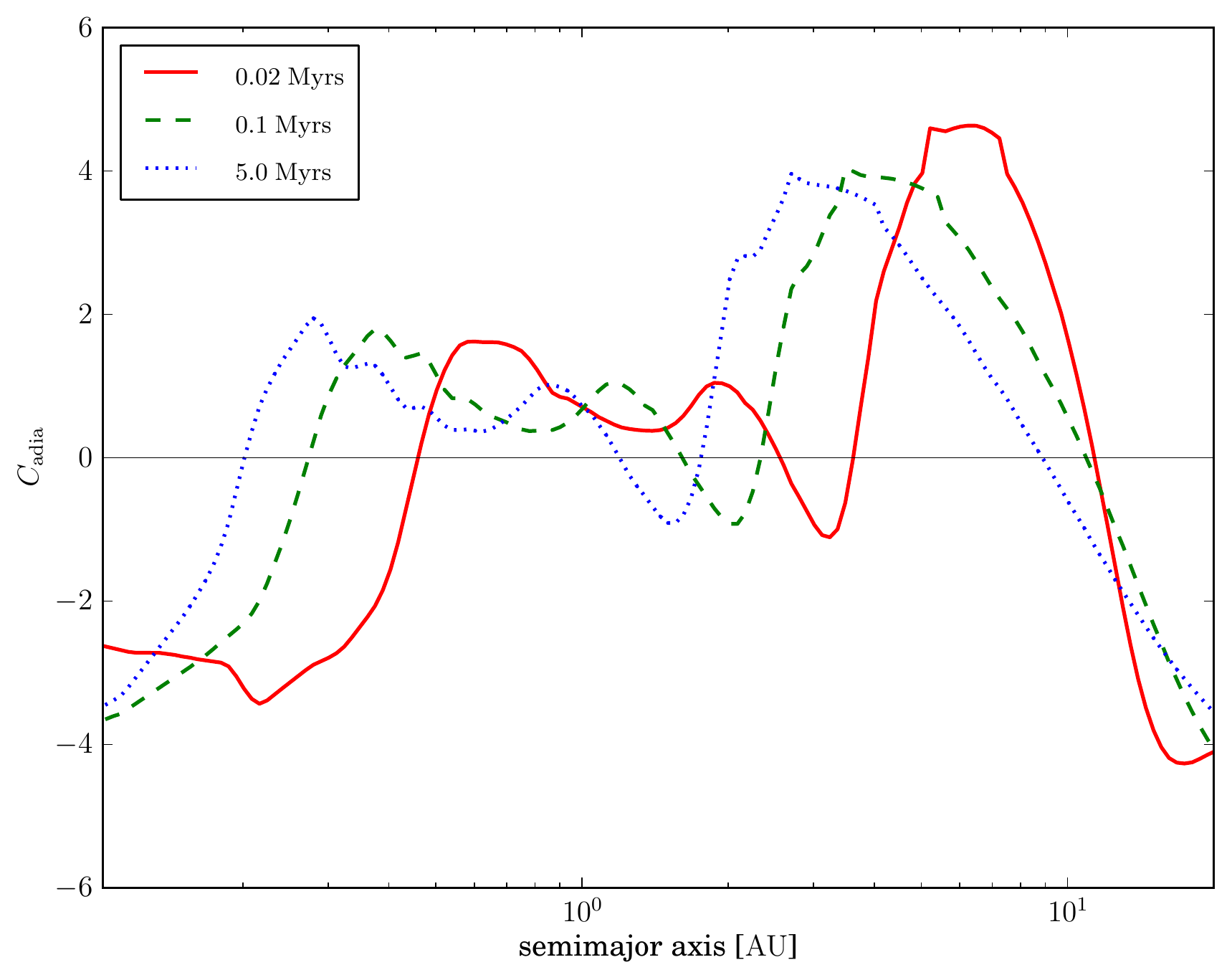} 
	\caption{Strength of adiabatic migration coefficient $\cadia$, plotted as a 
          function of   semimajor axis at times equal 0.02 , 0.1 and 0.5 Myr in a nonirradiated, evolving 
$\alpha$-disk. Positive values of $\cadia$ drive outward migration. }  
	\label{fig4} 
\end{figure}  
\begin{figure} 
        \centering \includegraphics[width=1.0\linewidth]{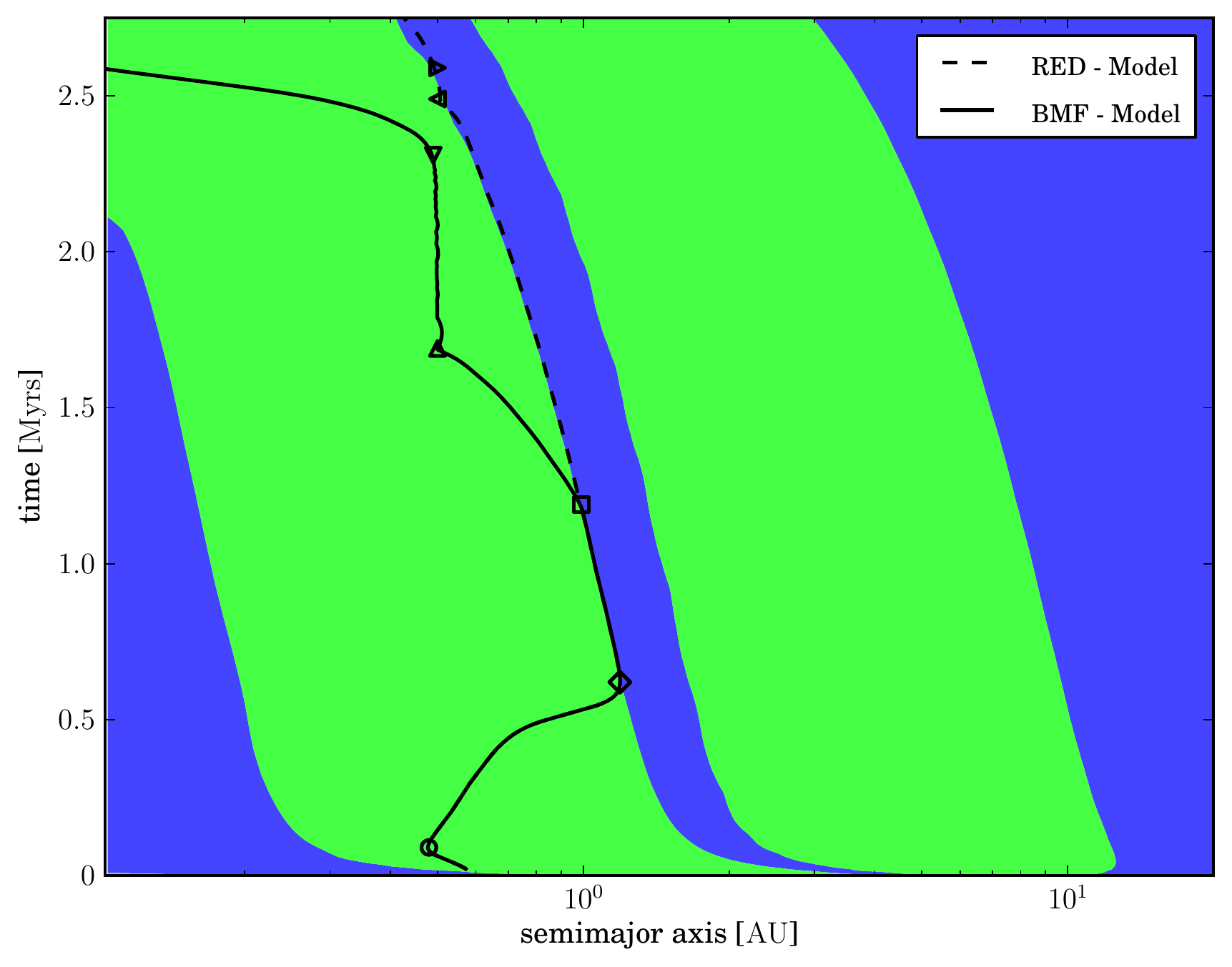} 
	\caption{Direction of migration in the adiabatic migration regime for 
          a nonirradiated evolving $\alpha$-disk. Blue indicates regions of 
          inward migration while green shows outward migration. The black lines are the 
          migration tracks of an evolving protoplanet set into the disk at ~2000 
          yr. The track shown with the solid line uses the BMF migration model while  
          the dashed line is calculated with the RED model (see Sect. \ref{migfortracks}). The symbols mark important points in the  evolution and are discussed in the text.}  
	\label{fig5} 
\end{figure}  
These positive torques lead to regions in a disk where a planet within the adiabatic migration regime migrates outward instead of inward \citep{lyrapaardekooper2010,mordasini2011}.
 Figure \ref{fig4} shows the values of $\cadia$ (which represents the strength of the migration in the adiabatic 
regime, see Eq. \ref{chap2 eq:cadia}) at time equal 0.02, 0.1 and 0.5 Myr in a nonirradiated evolving 
$\alpha$-disk with $\Sigma_0 = 420\ \rm{g/cm^2}$ and $\mwind = 6.7\times 10^{-9}\  
\msun/{\rm yr}$. These parameters results in a lifetime of the disk of 2.8 Myr. 

In Figure \ref{fig4} one can see two regions of outward migration, i.e., with positive values of $\cadia$, from 0.46 AU to 2.6 AU and farther out from 3.6 AU to 11.5 AU at 0.02 Myr (red curve). At later times these regions are closer to the star as 
the disk evolves. To further illustrate the time evolution of the regions in
Figure \ref{fig5}  we 
show the direction of migration as a function of time for this disk. 
Blue indicates regions of inward  migration while green shows outward migration. One can 
see that these regions slowly move inward as the disk evolves. This occurs on the viscous time scale of the 
disk \citep{lyrapaardekooper2010}. Owing to the existence of the outward and
inward regimes, there are special locations in the disk. For example, at an
age of 0.5 Myr, the migration changes from outward to inward at 10 AU (and at about 1.2
AU) when moving to a larger distance (i.e., these are points of zero torque
where the derivative of the torque with distance is additionally
negative). Such a point is called a convergence zone (CZ). It is called this
 because planets in its vicinity converge on this zone from both inside and
outside \citep[see][]{lyrapaardekooper2010}. The domain in orbital distances
from which planets migrate to the convergence zone is the associated
convergence region. At 0.5 Myr, the inner convergence region extends from 0.2
to 1.8 AU for
example. Similar results for two
convergence zones for various disk models were also recently presented by \citet{KretkeLin2012} 
and \citet{YamadaInaba2012}. After reaching the convergence zone, a planet
remains slightly outward, but close to it, so that the net torque pushes 
the planet inward at the same migration rate as the zone itself.

Once captured in a convergence zone, the planet migrates on a time scale that is at least an 
order of magnitude longer than typical type I migration time scales.  For
example, while captured in the inner convergence zone, the planet discussed
below has an equivalent migration coefficient $C$ (Eq. \ref{eqa}) of 0.034 while
$C$ is during normal type I migration on the order of 1 as shown for instance in Fig. \ref{fig4} 
or \citet{paardekooperbaruteau2010}. 
 
However, for some conditions, a low-mass planet cannot migrate at a sufficiently
high rate to remain close to the CZ. Instead, the planet leaves the CZ and falls behind
it (the planet still migrates inward, but is overtaken by the CZ). This occurs
especially for the inner convergence zone. Typically, after leaving the inner
zone, the planet transitions into the outer one, where it is again
captured. The reason is that the type I migration rate is proportional 
to the planet mass. For a sufficiently low mass of the planet, the type I
migration time scale thus becomes longer than the viscous time scale of the
disk, which sets the speed at which the CZ moves. This characteristically occurs at
the end of the disk lifetime, when the gas surface density is low, so that
the type I migration time scale becomes even longer.
 
The position of the inner CZ is close to the distance of the local minimum in 
the opacity at a temperature of $\approx 200K$  
\citep{lyrapaardekooper2010}, which is the temperature where ice grains are
completely molten in the opacity law of \citet{belllin1994}. 
The change in the slope of the opacity at this 
point leads to a change in the temperature power-law exponent $\betat$, which leads to the 
change of the sign of $\cadia$ and therefore to a convergence zone.  
The reason for the outer CZ is the change of the temperature slope due to the
convergence onto the 
background temperature in the outer part of the disk.  
 
The convergence zones only exist for a certain range of planet masses \citep{KretkeLin2012}. A low-mass planet will migrate in the locally isothermal regime since the thermal processes in the disk 
are fast enough to regulate the temperature during a u-turn of the gas. On 
the other hand, when a planets becomes massive enough, the angular momentum flux through the 
horseshoe region will be too low to support the unsaturated horseshoe drag, 
so this part of the torque saturates and in general rapid inward migration sets in.

\subsection{Migration and formation tracks} 
\label{migfortracks} 
\begin{figure} 
        \centering \includegraphics[width=1.0\linewidth]{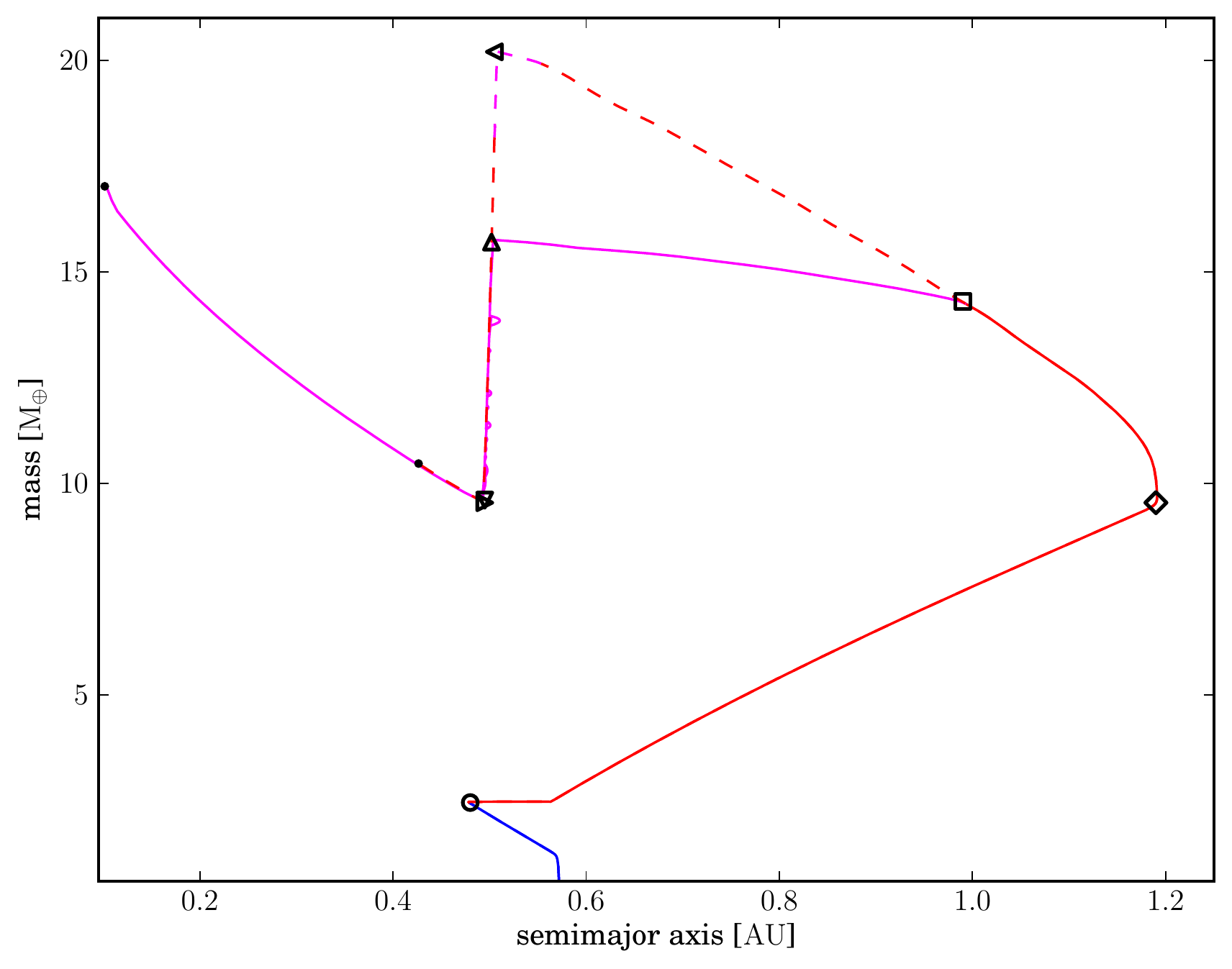} 
	\caption{Formation tracks, i.e., evolution of the position in the distance-mass  
          plane for a planet in a nonirradiated, evolving 
          $\alpha$-disk. The track shown as a solid line uses the BMF 
          migration model, while the dashed line uses the RED model. The colors represent 
          the different migration regimes and are also used in this way in 
          subsequent figures. Blue shows unsaturated locally 
          isothermal migration, while red shows unsaturated adiabatic and
          magenta saturated adiabatic migration. In both cases  
          the planet does not enter the type II or the saturated 
          locally isothermal regime. The symbols mark the same important points in the 
          evolution as in Fig. \ref{fig5} and are discussed in the text. The 
          small filled circles show the 
        final positions of the planets at the end of the simulations.} 
	\label{fig6} 
\end{figure}

The new migration model leads to changes in the evolution of a
protoplanet that are different from those described in \citet{mordasinialibert2009a}. We
discuss the new behavior in this section where we show the migration and
formation tracks of a typical protoplanet of our synthesis simulation. Not every evolving
protoplanet in our calculations shows every behavior described
below, but most evolutionary tracks of planets feature at least some part of
the behavior. The consequences of the new migration model for a whole
synthetic population are studied after this.

Further to what was discussed before, Figure \ref{fig5} also shows two tracks of an evolving protoplanet starting
at 0.57 AU at ~2000 yr after the start of disk evolution. Figure \ref{fig6}
shows the corresponding formation tracks of this protoplanet in  
the semimajor axis mass plane. In both diagrams, the simulation represented by the solid 
line uses the BMF migration model, where $\fvisc=0.55$,  while the dashed line 
is used for the RED model with $\fvisc = 0.125$ (see
Sect. \ref{modelversions}), simulating a larger saturation mass.   
The beginning of the evolution of the protoplanet is the same in both models.
 It starts to migrate inward in the locally isothermal 
migration regime (blue part of the track in Figure \ref{fig6}) and accretes solids from its surrounding, depleting this part of the disk of planetesimals. 
When it reaches a mass of approximately 2.5 $\mearth$, its horseshoe region is
broad enough that cooling cannot keep the gas that is on horseshoe  
orbits in a locally isothermal state. The planet enters the unsaturated adiabatic regime, shown in red in 
Figure \ref{fig6}, and starts to migrate outward (at the position marked by a 
circle in the figures). This occurs at 0.1 Myr as 
seen in the migration track in Figure \ref{fig5}. While migrating outward, the
planet does not grow at first because there is  almost no solid material left
because of its previous passage through this part of the disk. The protoplanet is  
also still too small to accrete significant amounts of gas. After crossing its 
initial position, growth by accretion of planetesimals sets back in and the
planet  migrates outward until it enters the convergence zone at 0.6 Myr
(diamond symbol).  Here the direction of migration again changes 
to inward as the planet is now bound to the CZ. While the disk evolves,
this zone and the captured planet move slowly inward. The planet again moves
through a region that is depleted of almost all planetesimals. But, with a 10 
$\mearth$ core, the planet is now massive enough to 
bind nebular gas in its envelope. This is especially true when there is not
much solid accretion, which means that the core luminosity is low, and
therefore the gas accretion rate is high. The planet grows by accreting gas,
until at about 1.25 Myr in the BMF model (solid line), it reaches 
the mass where saturation sets in (square symbol). The positive horseshoe drag
is reduced with increasing mass and is not sufficient to counterbalance the
negative Lindblad torques. The planets therefore leaves the CZ and rapidly  
migrates inward (saturated adiabatic migration regime, magenta in Fig. \ref{fig6}).  
 
The planet continues to accrete gas until the inner edge of its feeding zone
reaches the distance of the previous closest approach to the star (upward-facing
triangular symbol). Again in reach of planetesimals, solid accretion restarts
and increases the core luminosity. The envelope expands, heated by this process,
and because it is still connected to the disk at this time \citep[attached
phase, see][]{mordasini2012a}, the heating leads to the loss of envelope
mass and a corresponding reduction of the total planetary mass (see
Sect. \ref{accretionandinternalstructure}). This reduces the level of 
saturation, which in return increases the horseshoe drag, which can again balance the Lindblad torque resulting in an almost complete stop of migration: if the 
planet were to migrate outward again, solid accretion would stop and rapid gas 
accretion would start to increase the mass. This would reduce the horseshoe 
drag and push the planet back inward. In the opposite case of further inward 
migration, solid accretion would become stronger. This would increase the 
core luminosity and remove more envelope mass, and the total mass
decreases. This would increase the outward-directed horseshoe drag because of
the reduced saturation of the corotation torques. The planet would thus tend
to migrate outward. The combination of these two points
means that the planet has reached a quasi-stable point due to the interplay of
accretion and migration. 
   
With the evolution of the disk, the saturation mass at a given point in 
the disk decreases over time. Thus the mass where the partly saturated
horseshoe drag balances the Lindblad torques is also reduced with
time. Because of the
interaction of accretion and migration, the planet's mass remains exactly at
this zero-torque mass.

 The planet is forced to loose more mass and therefore remains at this
 semimajor axis for a given moment in time, while, as a result, the planet
 moves in slowly over longer time scales. It just ``nibbles'' on the edge of
 the planetesimal disk that was depleted up to this distance, while
reducing its mass. The gas envelope mass is decreased until a level is reached
that the planet can support with the full solid accretion rate of normal inward 
migration.  
 
In case of the BMF model, this is a slow process. The protoplanet remains for
0.5 Myr at 0.5 AU while it looses 6 $\mearth$ of the envelope it has accreted while moving 
inward in the CZ and later in the saturated adiabatic migration regime. At
this time it has only 0.1 $\mearth$ of envelope left. This is the same amount as at the time it 
became bound to the CZ. The RED model and its increase of the saturation mass leads to a slightly 
different behavior of the protoplanet. It remains for a longer time in the CZ and can thus 
accrete gas for a longer time than the BMF model protoplanet. This results in
a larger envelope mass when it reaches the distance of the previous 
closest approach (left-facing triangle). But since in both cases the
protoplanet migrated through the same part of the 
disk, both have the same amount of solids accreted and thus the same
core-mass. Therefore, the increase of the 
solid accretion rate also leads to a mass loss of the envelope in the RED model case. But here, the planet is 
still in the unsaturated migration regime while loosing most of its envelope mass. Therefore, the mass loss does 
not lead to a change in the migration rate. The protoplanet is still bound to the CZ and 
is pushed by it into the remaining planetesimal disk. Therefore, the mass loss occurs 
much faster here, the planets looses ~11 $\mearth$ in only ~0.1 Myr.  
 
This illustrates the strong inter-dependence of migration and accretion. The solid 
accretion rate is set by the amount of solids reachable by the planet, and 
therefore by the migration that brings the protoplanet into new regions of the 
disk. But this behavior is only true if the availability of planetesimals
themselves, and not the collision rate, is the limiting factor for the solid
accretion rate $\mdotcore$.  
In the planet-envelope structure model used here, the solid accretion
rate, the associated core luminosity, and the 
mass of the core itself define the envelope structure and thus the envelope mass of a 
planet. 

The small loops in the track that occur while the planet in the BMF model looses mass 
(between the up- and down- facing triangles) are caused by a finite 
time-step. Some larger loops are visible in the formation tracks in Figure \ref{figsyntracks}.  We 
show here the data obtained during the population synthesis calculations discussed in 
Sect. \ref{refsyncalc} and shown in that figure. We also separately simulated the same case with a much shorter time-step 
and obtained the same results without these small loops.  

We did not consider an increase in the random velocities of the 
planetesimals due to the presence of a planet. They remain small, as reported in 
\citet{pollackhubickyj1996}. A more realistic oligarchic growth model as
described in \citet{fortierbenvenuto2007} 
and \citet{Fortier2013} leads to higher random velocities and thus a smaller focusing 
factor in the calculation of the solid accretion rate $\mdotcore$. This would 
lead to less envelope loss if $\mdotcore$ remains small enough.   
 
Another point to consider, in addition to the planetesimal accretion rate, is the treatment of the protoplanetary disk in our 
model. We did not let the disk evolve except for depletion due to accretion onto 
the protoplanet. Therefore there is a sharp edge into which the planet can 
migrate. A more realistic treatment (e.g., planetesimal drifting or diffusion, 
scattering) would lead to a gradual increase of 
$\mdotcore$ and thus to a slower loss of the envelope. But we expect the final
outcome to be similar, only the track in the a-M plane would be smoother (mass
loss setting in earlier and more gradually).

In both cases shown here, the protoplanet starts again to accrete solids and 
nebula gas and migrates inward after it looses almost all its envelope 
mass (down- respectively right-facing triangles in Fig. \ref{fig5} and
\ref{fig6}). In case of the BMF model it migrates in the faster saturated 
adiabatic migration regime until in reaches 0.1 AU and a final mass of ~17 
$\mearth$ (~0.9 $\mearth$ in its envelope) and the simulation stops.   
In the RED model the simulation ends with the disappearance of the disk at 2.8 
Myr and the planet migrated to 0.43 AU with a mass of ~10.5 $\mearth$ (only 
0.1 $\mearth$ in the envelope).

\subsection{Reference population synthesis calculation} 
\label{refsyncalc} 
\begin{figure} 
        \centering \includegraphics[width=1.0\linewidth]{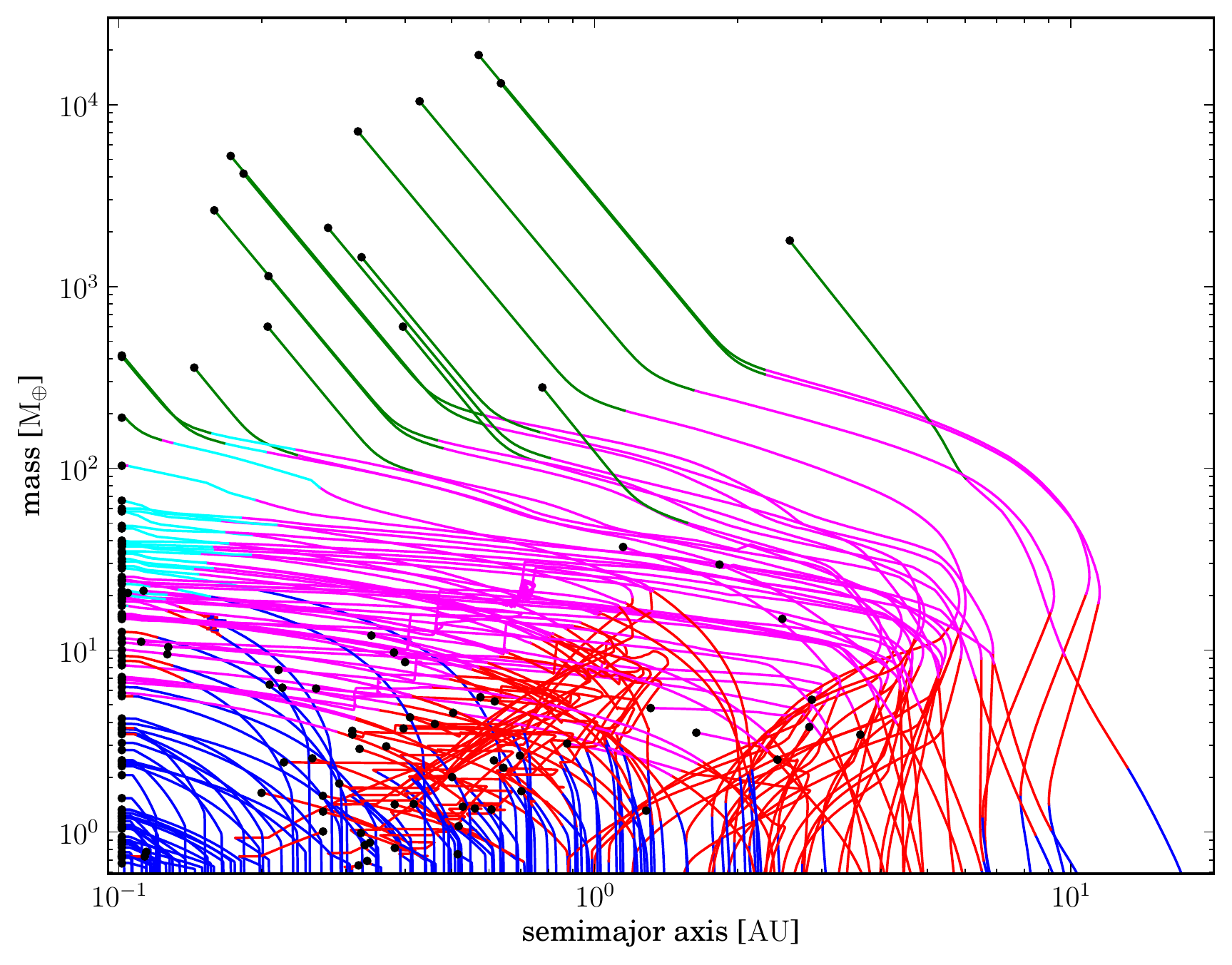} 
	\caption{Formation tracks, the evolution of the position of the planets in 
          the semimajor axis mass diagram. Color shows the 
          different migration regimes at this point of its formation. Blue shows unsaturated locally 
          isothermal, cyan saturated locally isothermal migration, red shows 
          unsaturated adiabatic and magenta saturated adiabatic migration, 
          finally, green shows type II migration. The filled circles show the 
        final positions of the planets at the end of the simulations. } 
	\label{figsyntracks} 
\end{figure} 
 
\begin{table} 
 
\begin{center} 
\begin{tabular}{l c} 
\hline \hline 
Quantity & value  \\ \hline 
Initial disk power-law exponent          & -1.5 \\ 
Disk viscosity parameter $\alpha$ & $7\times 10^{-3}$\\ 
Inner radius of computational disk & 0.1 AU \\ 
Outer radius of computational disk & 50 AU \\ 
Gas surface density at inner radius & continues \\ 
Irradiation for disk temperature profile & not included \\ 
Iceline & included \\  \hline 
Embryo starting mass & 0.6 $\mearth$ \\ 
Core density & $3.2 \rm{g/cm^2}$ \\ 
Envelope type & primordial H$_{2}$/He\\ 
$dl/dr$ in the envelope & zero \\ 
Grain opacity reduction factor & 1.0 \\ 
Type I migration & BMF-model \\ 
Type I migration reduction factor & none \\ 
Transition criterion type I to type II & \citet{cridamorbidelli2006}\\  
Transition exponent \\ Type I to type II migration (Eq. \ref{cfinal})& 10.0 \\ 
Transition exponent \\ Unsat. to saturated migration (Eq. \ref{ct1v2}) & 4.0 \\ 
Cooling reduction factor $\fcool$ & 1.0 \\ 
Viscosity reduction factor $\fvisc$ & 0.55 \\ \hline 
Stellar mass & 1 $\msun$\\ 
Simulation duration & till gas disk vanishes \\  
Number of embryos per disk & 1 \\\hline 
\end{tabular} 
\end{center} 
\caption{Parameters and settings for the reference population synthesis. }\label{tab:popsynth} 
\end{table}%

After studying the single case, we now look at a population synthesis 
calculation with 10000 different initial conditions that we used as our
reference when we investigated the effects of different migration models on a
synthetic planet population. 
The important parameters of the synthesis can be found in Table \ref{tab:popsynth}. 
We used the BMF-migration model and an $\alpha$ parameter of $7\times 10^{-3}$ for 
the nonirradiated disk.  
 
Figure \ref{figsyntracks} shows the tracks of 250 cases, the different
migration regimes color coded. The meaning of the colors is described in the
caption of Fig. \ref{figsyntracks}.  
One can see that most planets start in the locally isothermal migration regime, changing 
into the unsaturated adiabatic migration regime before the horseshoe drag saturates and 
they migrate inward. Some are large enough to transition into type II 
migration while others end up at 0.1 AU. Two groups corresponding to the inner 
(inside of 1AU) and outer (outside of 1 AU) convergence zone are visible in Fig. \ref{figsyntracks}.  
  
\begin{figure} 
        \centering \includegraphics[width=1.0\linewidth]{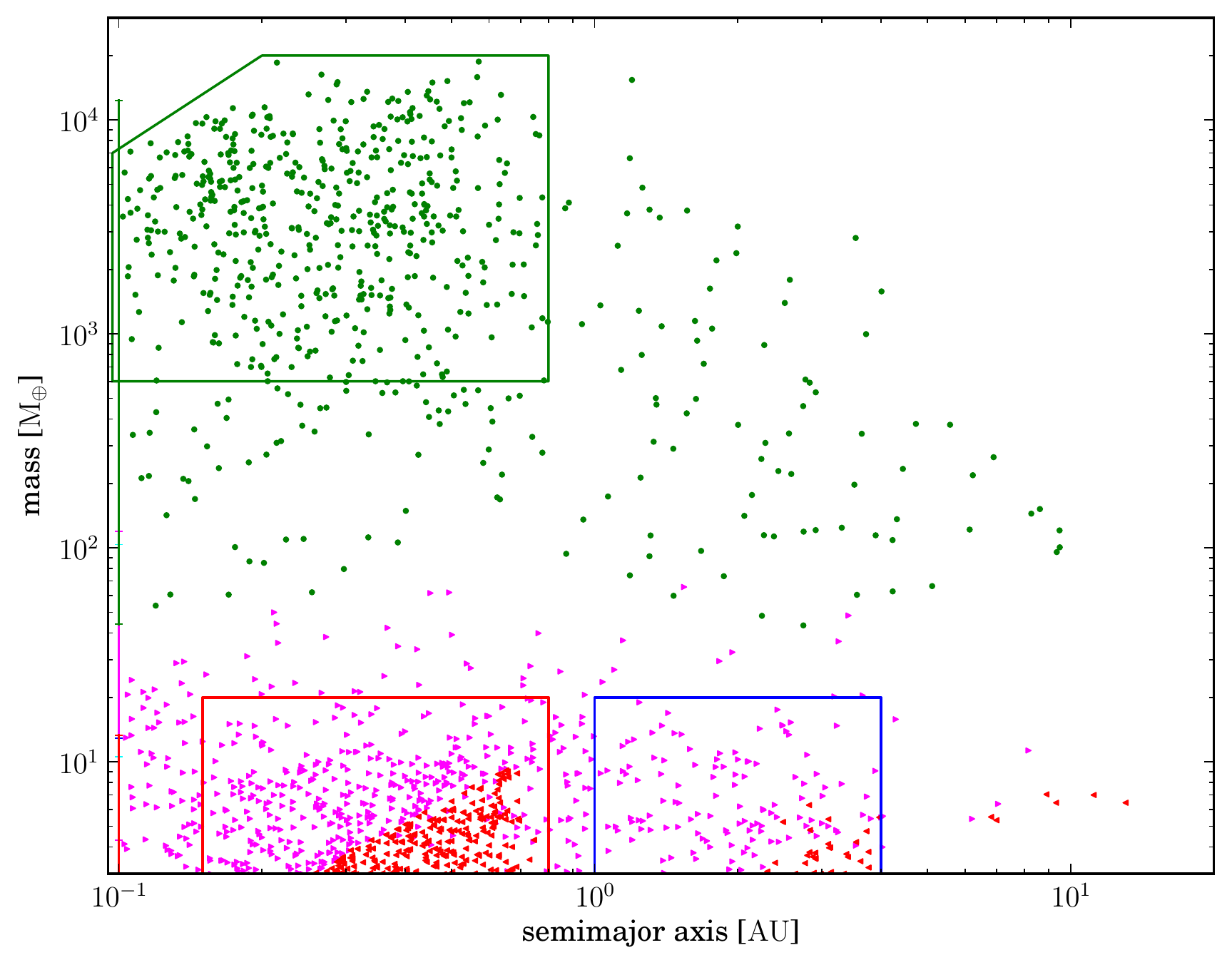} 
	\caption{Final position of the synthetic planets in the semimajor axis mass 
          diagram. Color shows the different migration regimes a planet is in 
          when the calculation ended. The 
          meaning of the colors is the same as in Fig. \ref{figsyntracks}.  The bars at 0.1 AU indicate 
          the mass range of the ``hot'' planets in the different regimes. The 
          boxes indicate clusters of planets described in the text.}  
	\label{figsynAM} 
\end{figure}

When a planet migrates outward through the iceline, the migration rate does not change, 
but its solid accretion rate does change by a factor of 4, because the increase 
in the solid surface density. This results in a 
bend and a much steeper slope in some of the tracks in the outer
zone. Horizontal parts in the formation tracks that indicate migration without
growth are also lacking in the tracks of the  outer planets. In the outer parts,
especially outside of the iceline, the amount of solid matter is too large to
be completely accreted onto the planet at the given accretion rates
$\mdotcore$. Thus, planets can also collect material on their second pass
through a part of a disk and grow.   
 
The inner and outer CZ are the reason for three groups of planets in the final position 
a-M diagram. They can be seen in Figure \ref{figsynAM}. A fourth possible
cluster is not visible because the BMF-migration model leads to a situation
where all those planets migrate inward of 0.1 AU (see Section \ref{satmasssyn}):  
\begin{itemize} 
\item The first cluster, which is the one with the highest number of planets, 
  lies approximately between 0.15 
  and 0.8 AU and ~1 and ~20 $\mearth$ (red box in Figure \ref{figsynAM}). It consists of planets captured by the 
  inner convergence zone. Most of the planets are directly attached to the CZ when the disk 
  disappears and the simulation stops. Their mass is too small for a 
  transition into saturated migration and departure of the CZ. A strip of
  planets extends inward from this cluster. These planets saturated  
  shortly before the disk ended. However, most planets that transition into the 
  saturated migration regime while being in the inner CZ eventually end up at 0.1 AU.  
\item The second cluster lies farther away from the star (1 AU to 4 AU) but is also in the mass range 
  between ~1 and ~20 $\mearth$ (blue box in Figure \ref{figsynAM}). It consists of planets captured by the outer 
  second CZ. It is less populated because fewer embryos start at the larger 
  distances because the distribution of the starting positions, which are 
  uniform in log($a$). Additionally, the amount of solid material to grow is 
  larger, thus many planets at these distances can reach masses above the saturation mass.   
\item The third cluster consists of planets that saturated 
  while being in the outer convergence region (green box in Figure
  \ref{figsynAM}). But they are massive enough and can accrete  enough solids
  on their way in that most of them can transition into type II  
  migration. Here they migrate on the time scale of the disk evolution or 
  slower and can accrete gas until the disk ends. They reach masses between 100 
  and a few 10000 $\mearth$. The planets can reach such high masses because we 
  neglected the effect of gap formation on the gas  accretion rate. If the reduction of the gas accretion rate due to gap formation were included, the planet masses would be restricted to lower masses, depending on the disk viscosity and mass \citep[cf.][]{bodenheimer2013}.
\end{itemize} 

Planets in the inner convergence region will  
migrate to the inner CZ and thus migrate through a large part of the inner part of the 
disk. But they remain completely inward of the iceline since the inner
convergence region ends there due to opacity transitions. Therefore they are
able to accrete most matter in the first few 0.1 AU and thus finally have at
least a few Earth masses. In our model we obtain a small planet of several
$\mearth$ or less only when its starting time is in the last few 0.1 Myr of
the disk lifetime. With more than one core per disk one planet alone cannot accrete
all the matter in the inner part. The solids will be distributed in many
small planets. We therefore overestimate the number of planets between 2 and
30 $\mearth$ and underestimate the number of planets smaller than 2 $\mearth$.

The synthesis with the new migration model also shows the desert of 
planets between 30 and 200 $\mearth$. This is a feature of the runaway gas 
accretion that occurrs in the core-accretion model \citep{pollackhubickyj1996}.   
On the other hand, the region of close-in, low-mass planets, which was empty in
\citet{mordasinialibert2009b}, is now well populated with the new
nonisothermal migration model.  
 
The spread of the first cluster originates in the spread of the initial 
conditions of the photoevaporation rate $\dot{\Sigma}_{\rm w}(a)$. In the implementation of 
our disk module the photoevaporation rate determines the mass of 
the disk at the end of a simulation and therefore the position of the CZ at 
the end of a simulation. A single value of $\dot{\Sigma}_{\rm w}(a)$ in all simulations of a 
synthesis would result in only one position of the CZ and therefore a high 
concentration of small mass planets on one radius (see also Sect. \ref{sect:convzone}).   
 
We used some basic statistics, namely the number of ``hot'' and ``cold'' planets and the 
number of massive and small planets (cf. below), to compare synthesis calculations 
with different migration models. While none of these numbers are compared with 
those of the  observed population of extra-solar planets, the difference between 
the calculations allows us to see the importance of different 
parameters or parts of the migration model. Comparison with the observed population will be made in future work  
when other such as effects like the decrease of the disk mass due to accretion
onto the planet or multiple concurrently forming planets
\citep{mordasini2012a,mordasini2012b,Alibert2013} are also  
included. Out of 10000 initial conditions we obtained ~6850 planets more massive than our starting mass of 0.6 
$\mearth$ in the synthesis calculation 
described above. The remaining ~3150 initial conditions have starting times (time
when we insert the planet embryo into the disk) longer than the lifetime of the corresponding disk. 
Out of these 6850 planets 54,4\% migrated to 0.1 AU, the inner border of the 
computational disk, and are
called ``hot'' planets. The other 45,6\%  ended further out and will be called ``cold'' 
planet.  
 
Finally, there are 912 (13.3\%) massive planets in total with $M > 100 \mearth$. While 
the majority of all the planets in the synthesis are ``hot'', the massive planets 
split into 276 ``hot'' and 636 ``cold'' planets. Thus there are more ``cold'' 
giant planets than ``hot'' ones.  
 
\section{Impact of different migration prescriptions} 
In this section we compare results from population synthesis
calculations where we changed one aspect of the migration model relative to the calculation above
(Sect. \ref{refsyncalc}), but otherwise used the same initial conditions and
settings. We therefore refer to the synthesis above as the reference synthesis.
We first compare it with migration models described in earlier studies before we change some physics
of the model itself.

\begin{table*} 
\centering 
\begin{tabular}{|l|c|c|c|c|c|c|} 
\hline 

name & ``hot'' & ``cold'' & total massive & ``hot'' massive & ``cold'' massive 
\\ \hline 
BMF model, reference synthesis & 54.4 & 45.6 & 13.3 & 4.0 & 9.3 \\  
isothermal migration model           & 79.2 & 20.8 & 7.8 & 2.7 & 5.1 \\  
BMF model, Casoli Lind.        & 48.8 & 51.2 & 15.0 & 2.9 & 12.1 \\  
RED model                     & 35.4 & 64.6 & 19.1 & 1.4 & 17.7 \\  
STD model                     & 60.5 & 39.5 & 11.4 & 4.5 & 6.9 \\  
Paardekooper, free gamma      & 55.4 & 44.6 & 12.0 & 4.1 & 7.9 \\  
Paardekooper, gamma = 1.4     & 59.3 & 40.7 & 9.5 & 4.1 & 5.4\\  
BMF model, irradiated disks   & 68.8 & 31.2 & 10.1 & 3.8 & 6.3\\ \hline 
 
\end{tabular} 
\caption{Statistical results of population synthesis calculations. In the 
  first seven syntheses simulation we consider 6850 planets more massive than 
  $0.6\ \mearth$. In the synthesis with the irradiated disk we consider 
  $\approx 7700$ planets more massive than $0.6\ \mearth$. The second column shows the percentage of planets that 
  migrated to 0.1 AU (``hot'' planets), while the third column corresponds to 
  ``cold'' planets ($a > 0.1 \rm{AU}$).We also show the fraction of embryos that
  grow more massive than $100\ \mearth$ (total massive) and how they split 
  into ``hot'' and ``cold'' massive planets in columns four to six.} 
\label{tabsyn} 
\end{table*} 

\subsection{Earlier prescriptions}

We use two earlier prescriptions, the isothermal migration model of \citet{tanakatakeuchi2002} used in our earlier work
and the migration prescription from \citet{paardekooper2011}.
 
\subsubsection{Isothermal migration model of Tanaka et al. 2002 } 
\label{oldmodel} 
\begin{figure} 
        \centering \includegraphics[width=1.0\linewidth]{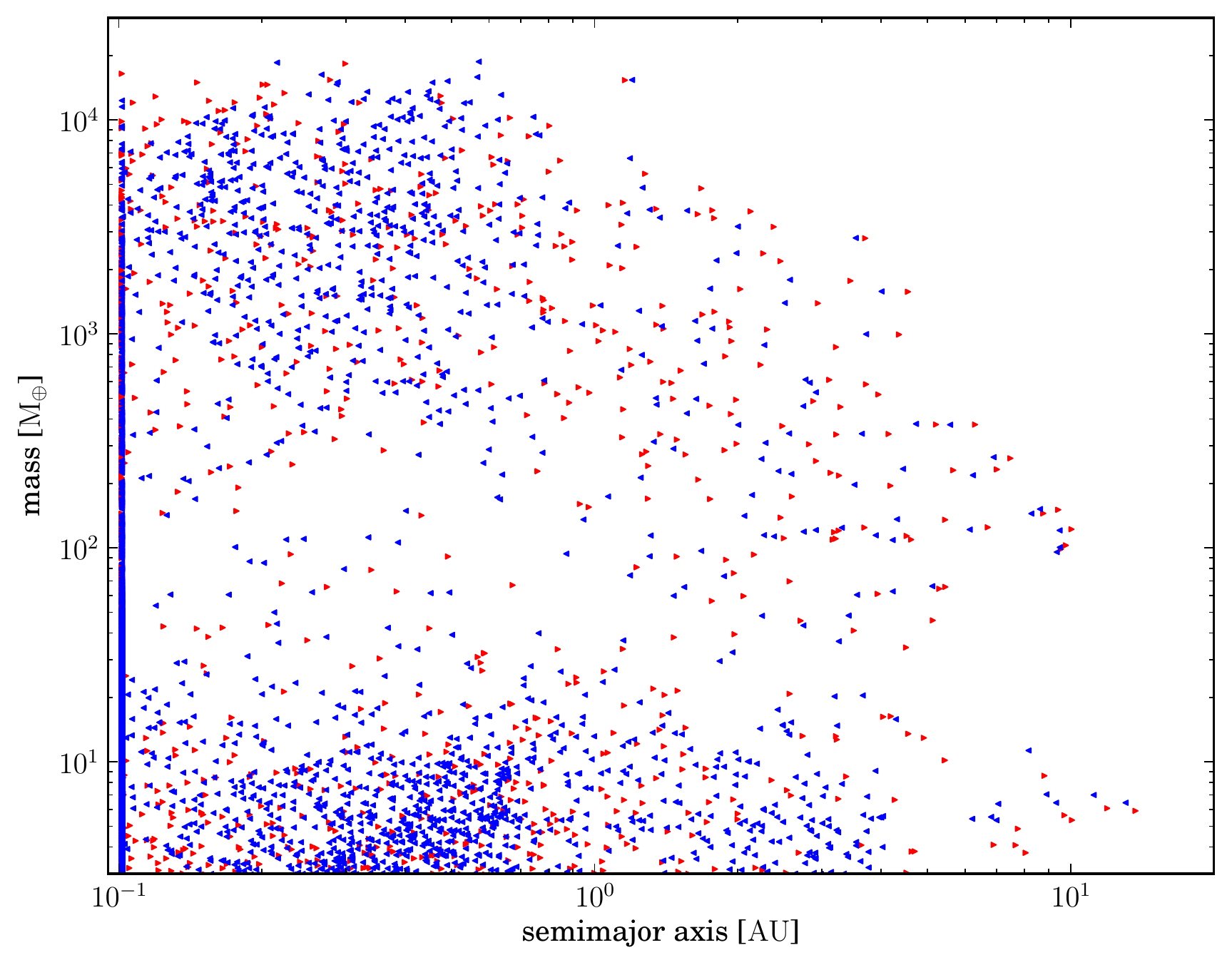} 
	\caption{Final position of the planets in the distance - mass 
          diagram. Red, right-facing triangles shows the positions obtained with 
          the isothermal migration model. Blue, left-facing triangles 
          correspond to the reference synthesis.} 
	\label{oldmig_AM} 
\end{figure} 
 
The first comparison was made with calculations preformed with the nonreduced isothermal migration model 
($f_1 = 1$) based on the results of \citet{tanakatakeuchi2002}. While we used
the same type I migration model as in \citet{mordasinialibert2009a}, the results here still 
differ from those published in \citet{mordasinialibert2009a}, since we here
use the same transition criteria for the transition from type I into type II migration
as in the reference synthesis. This is different from the one in
\citet{mordasinialibert2009a}. There only the thermal condition ($H > \rhill$) for the transition into type II was used. This leads to much smaller transition masses than here.

Figure \ref{oldmig_AM} shows the 
final position of the planets in the distance-mass diagram in red,
right-facing triangles. Blue, left-facing triangles show the reference 
synthesis. While the range in mass and distance covered by the planets is the
same, there is no clustering caused by the migration into a CZ at small masses (0.6 to 30 $\mearth$). 
The total number of ``cold'' planets ($a > 0.1\ AU$) is only $\approx 45\%$ of 
the number found in the reference calculation. From the 6850 synthetic planets more 
massive than 0.6 $\mearth$ we find 79.2\% ``hot'' planets  with  2.7\% ``hot'' massive planets 
and only 20.8 \% planets outside of 0.1 AU. Of these, 346 are more massive than 
100 $\mearth$ (5 \%). The new nominal nonisothermal migration model in comparison gives twice as many 
planets that remain outside of 0.1 AU and also almost twice as many massive 
planets. Some preliminary calculations indicate that this ratio can increase even more 
with lower values of $\alpha$. The nominal migration model still 
leads to a loss of more than half of the planets with more than 0.6 $\mearth$ into 
the inner part of the disk inside of 0.1 AU and therefore potentially into the 
star. On the other hand, the new model doubles the number of planets outside of 0.1 
AU compared with the isothermal migration model without any arbitrary 
reduction factor.

\subsubsection{Paardekooper et al. 2011 migration model} 
\label{paardepopsyn} 
\begin{figure} 
        \centering \includegraphics[width=1.0\linewidth]{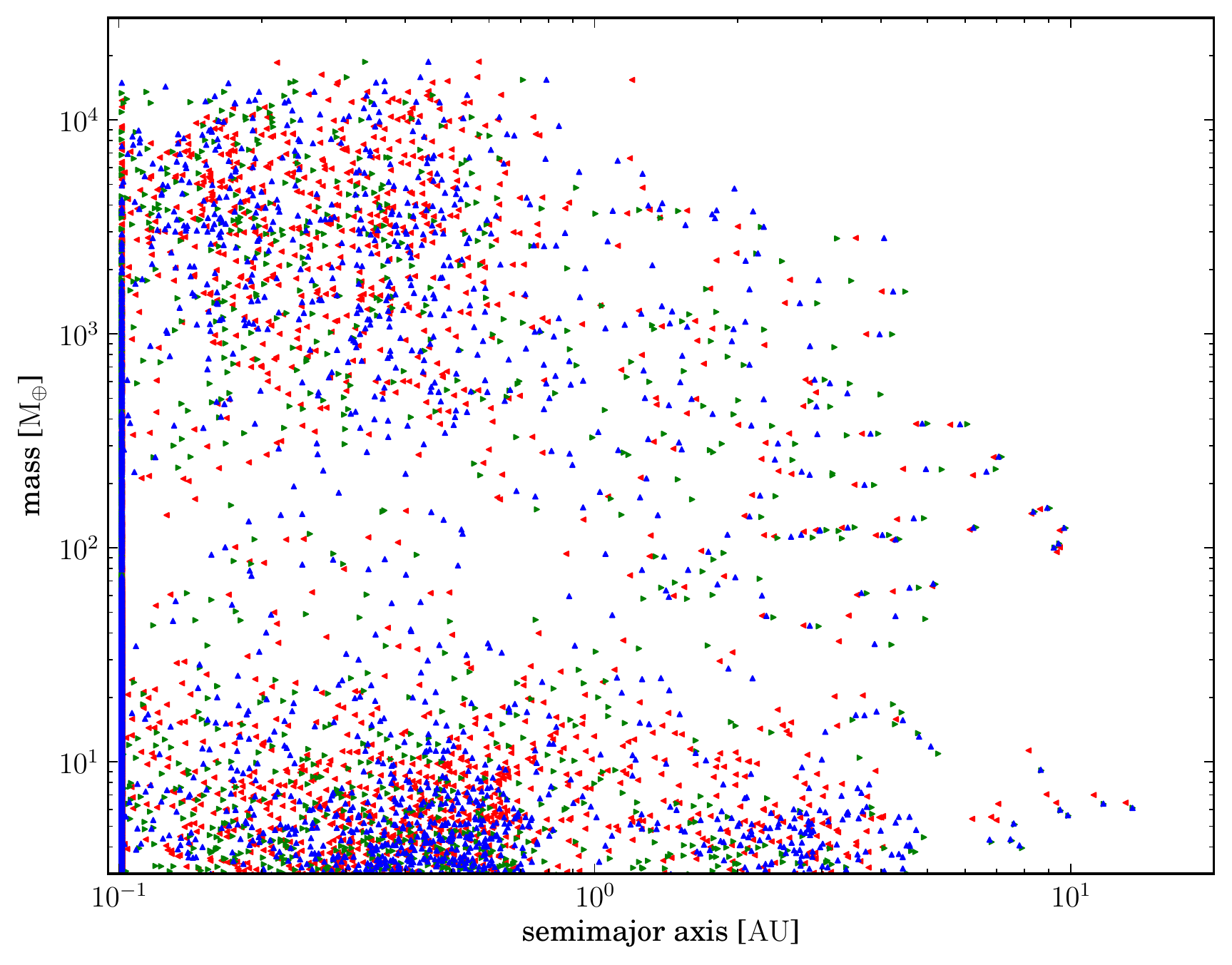} 
	\caption{Final position of the planets in the distance-mass 
          diagram. Three different syntheses are shown. Blue, up-facing triangles show the 
          reference synthesis. The other two population 
          calculations are made with the migration model of Paardekooper et 
          al. (2011). Red, left-facing triangles represent planets obtained with this model and the 
          adiabatic coefficient $\gamma$ calculated by our EOS. Green, right-facing triangles show 
          the a-M positions obtained with the Paardekooper et al. 2011 model with a fixed $\gamma =1.4$.} 
	\label{paarAM} 
\end{figure} 
 
\citet{paardekooper2011} developed a migration model that is similar to the one described in Section \ref{migmodel}, but more sophisticated, because it uses thermal-diffusion time scales and viscous time scales as transition criteria between
barotropic and entropy-related parts of the horseshoe drag and the saturation of 
both. The difference between their and our model are discussed in Section 
\ref{compmodel}, where we compared torque curves from our model with 
those of this model. We also made two synthesis calculations using this 
migration model. The final a-M distribution of the two simulations is shown in 
Figure \ref{paarAM}. Green dots represent the final positions of the synthetic 
planets obtained with the \citet{paardekooper2011} model and a fixed $\gamma=1.4$. The second calculation is shown in red, here $\gamma$ was determined 
with the EOS we usually use \citep{SaumonChabrier1995}. The blue symbols show the reference synthesis.  
 
For planet masses larger than three Earth masses there is no real difference 
between the three plotted syntheses. The simulation with the free $\gamma$ produces 
nearly the same number of ``cold'' planets, the simulation with the fixed
$\gamma$ slightly $(5-10\%)$ less relative to the nominal BMF model. The
situation is slightly different for the number of massive planets with $\mplanet
> 100 \mearth$ outside of 1 AU: the free $\gamma$ synthesis has only $85\%$ of
the number of ``cold'' massive planets of the BMF-model synthesis  and the
fixed-gamma simulation about $60\%$ of the nominal model.  
 
The reason for the small differences in the overall amount of ``cold'' planets but the larger differences for the ``cold'' massive planets is, as visible in Fig. \ref{fig1} and \ref{fig2}, 
 that overall both models produce the same general migration behavior: first planets migrate inward, then outward to a convergence zone, and after saturation inwards again.
But the point of crossover from outward to inward migration is closer in and at lower masses 
for the \citet{paardekooper2011} model relative to the nominal BMF model. The planets saturate at lower 
masses and therefore start to rapidly migrate inward earlier in their 
evolution and fewer planets can reach the type II migration regime and become 
massive.  

Overall, there are no large difference between the different migration models.  
On the other hand, our simpler BMF model seems to agree better with 
the torques obtained with the radiative hydrodynamical simulations of 
\cite{kleybitsch2009} and \cite{bitschkley2011}.  

\subsection{Different input physics}
 
We now study the effect of different Lindblad torques and of different
saturation masses onto the synthetic planet population.

\subsubsection{Lindblad torques} 
\label{sec:Lindblad Torques} 
 
\begin{figure} 
        \centering \includegraphics[width=1.0\linewidth]{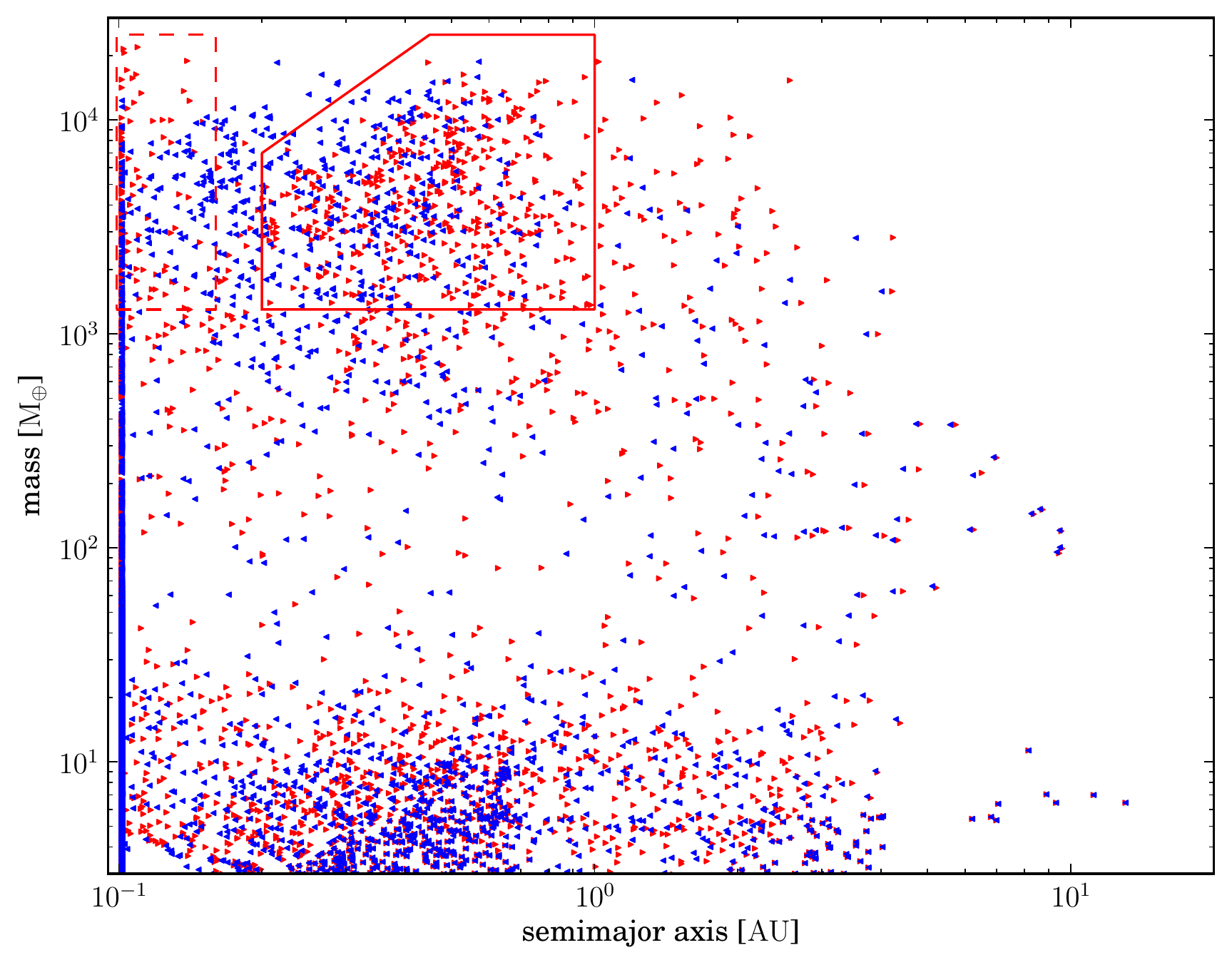} 
	\caption{Final position of the planets in the semimajor axis mass 
          diagram. Red shows the final position of planets obtained with the 
          BMF model but with a Lindblad torque formula from  \citet{massetcasoli2010}. 
          In blue are depicted the results of the reference synthesis.} 
	\label{lindAM} 
\end{figure} 
 
As stated in Section \ref{theory}, there are two different formulas for the 
Lindblad torques. We changed the Lindblad torque to the one 
used in \citet{massetcasoli2010} for one synthesis. We show the final positions of the 
planets in the distance-mass diagram in Figure \ref{lindAM}. The color-coding is 
the same as before.   
The resulting Lindblad torque is weaker with the equation of \citet{massetcasoli2010}, thus the migration in the saturated 
regime is slower. This results in moving the high-mass, third clusters farther 
out. In the reference synthesis the outer cluster is located between 0.1 AU and 
0.6 AU. Here, it is located between 0.2 AU and 1 AU (red solid box). Inside of 0.2 AU one 
can see some planets of a fourth cluster. The planets in this inner group
originate in part from fast growing planets, that is, those large dust-to-gas-ratios, of the inner convergence zone. The rest are planets from the outer zone, which saturated but grow rapidly enough during their fast inward-migration phase to transition into type II migration. In contrast to the 
reference synthesis, where all planets of this group migrated to the inner border of the computational domain at 0.1 AU, the planets are now at larger distances and can be seen in the 
calculation (red dashed box). The third cluster (outside of 1AU) completely
consists of planets in the outer convergence zone.

The smaller torque also results in a smaller and weaker region of inward migration between the CZ. It is small and weak enough that in 
some cases, planets can drift from the inner into the outer convergence 
region, because the migration rate is lower than the movement rate of the 
CZ due to the evolution of the disk (see Sect. \ref{sect:convzone}). 
  
For low-mass planets below the saturation mass the differences are 
smaller between the two syntheses. There are still the two clusters, associated with the two convergence 
zones. 
 
We can thus conclude that the weaker Lindblad torque results in an shift of the 
final position of a planet in the distance-mass diagram to the 
right that is, to larger distances. Because of this we also see more ``cold''  
(increase from 45.6\% to 51.2\%) and more massive planets relative to the 
reference synthesis (increase from 9.3\% to 12.1\%). The calculated migration
rate in saturation is only about a factor of 2-3 smaller with the formula of
\mbox{\citet{massetcasoli2010}} than that of \citet{paardekooperbaruteau2010}. This
seems small compared with the previous 
reduction of type I migration by a factor of 10-1000. But still, this small change in the 
description of a part of the torque by a factor of three has observable 
influence on the distribution of massive planets by up to a factor of 2 in 
semimajor axis for some cases.  
 
\subsubsection{Saturation mass} 
\label{satmasssyn} 
\begin{figure} 
        \centering \includegraphics[width=1.0\linewidth]{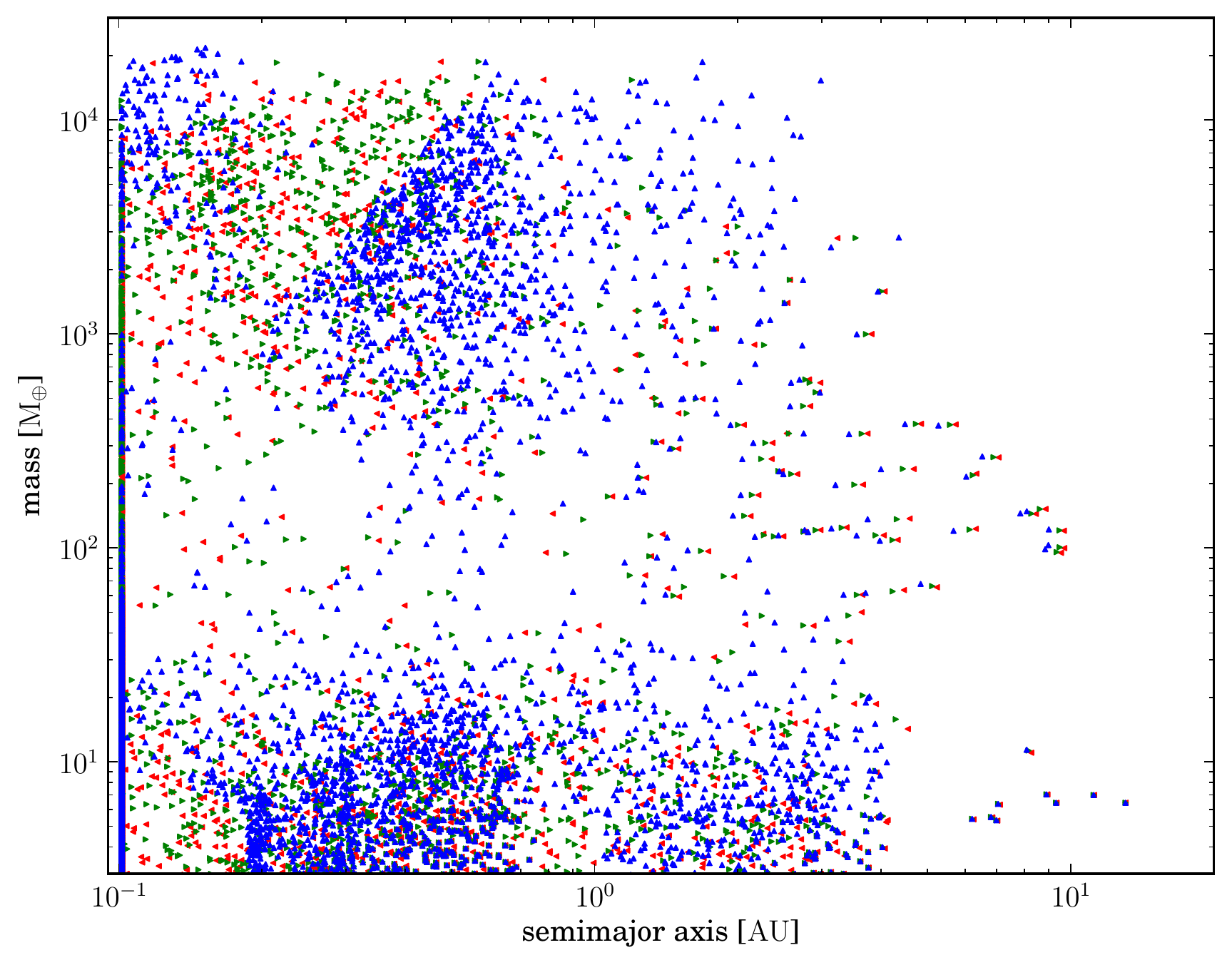} 
        \caption{Final position of the planets in the semimajor axis mass 
          diagram. Shown are three different synthesis calculations with different 
          values of $\fvisc$ and therefore three different saturation mass levels. 
          Red shows the calculation with the largest $\fvisc$, the STD case. 
          In green is shown the BMF case, and in blue the RED case with the 
          smallest $\fvisc$.} 
	\label{satmassAM} 
\end{figure} 
 
We furthermore conducted population synthesis calculations with two different values of $\fvisc$. One with 
a larger $\fvisc = 1$ (STD model) and one with a smaller $\fvisc=0.125$ (RED 
model) than the reference synthesis (BMF model, $\fvisc=0.55$) (see also \ref{modelversions}). 
This only affects planets that are massive enough to undergo the transition into a 
saturated migration regime. An eight times smaller $\fvisc$ will result in a four times 
larger saturation mass (Eq. \ref{eqmsat}). Figure \ref{satmassAM} shows the 
final positions of the planets in the distance-mass diagram for the two nonnominal 
calculations and for the reference synthesis. The blue (red) points are the  
RED (STD) case. For most points of the two clusters of low-mass planets there 
are only small differences in between these three simulations. They result from the onset of saturation at 
different masses. For example, some of the planets in these clusters will start to saturate 
with the STD model while they are still in the unsaturated adiabatic migration 
regime at the same mass with the RED model and thus migrate faster and also accrete differently 
because of the different migration rates. 
 
 The high-mass clusters are shifted farther out in the synthesis with 
 $\fvisc=0.125$ than in the reference calculation. Moreover, the fourth group 
 mentioned in Sect. \ref{sec:Lindblad Torques} is visible for planets obtained
 from the RED case. In this situation the time 
 planets spend in the saturated migration regime is shorter because of the 
 higher saturation mass, therefore the distance they migrate inward is 
 smaller and the planets end up farther out. This has an effect on the 
 content of heavy elements in the planet. Less migration means that the planet can reach less amount of 
 planetesimals in our disk. The difference is the amount of solids in the
 annulus of the planetesimal disk that is not visited by the planet with the larger saturation mass. But the difference in the final mass for 
 most massive planets ( $> 1000\ \mearth$) is small, lower than $1\%$. The reason is that they 
 grow the most while they are in type II migration and the runaway accretion phase. In this phase the accretion is 
dominated by gas accretion. Here the growth is set by the 
remaining disk lifetime, and this is almost the same for the different 
cases.

 Note that we made one simplifying assumption: we included the eccentric 
instability \citep{KleyDirksen2006}.  Therefore gap formation does not lead to a 
reduction of the planetary gas accretion rate. If this effect were not 
included, the maximal planet masses would be on the order of 10 Jovian masses \citep{Lubow1999,Armitage2007}.  
 
Planets beyond 2 AU and between 60 and $400\ \mearth$ are closer to the 
star with a higher saturation mass level than the planets mentioned 
above. The disk temperature is lower in the outer parts and the 
slope of the temperature profile begins to become flatter ($\betat \rightarrow 0$) as 
the temperature approaches the assumed background temperature of $10\ \rm 
K$. This change in the slope leads to a strong change in the horseshoe drag at a radius 
around 10 AU for early disk times and farther in at later times. It dictates the 
position of the outer convergence zone \citep{KretkeLin2012}. The change 
of the slope causes the horseshoe drag itself to become negative not far 
outside of the outer CZ and pushes the 
planet inward, as does the Lindblad torque. This means that an increase in 
the saturation mass will result in faster overall inward migration for the planets that start 
outside the outer CZ and are saturated in this part of the disk. 
 
When comparing the numbers of ``cold'' planets in the STD and the RED case we see that the 
increase of the saturation mass by a factor of 4 results in 64\% more ``cold'' 
planets and also 156\% more ``cold'' massive planets. The number of ``hot'' massive 
planets is reduced from 7.5\% to 1.4\% as the slower overall migration shifts the 
clusters outward. Even if we cannot directly compare our results with
observations, as a reference, \citet{Mayor2011} stated an observational value
of $\approx 1\%$.   
 
\section{Irradiated vs nonirradiated disks} 
We recently included irradiation of the host star into the disk model assuming an equilibrium flaring angle
\citep{fouchetalibert2012}. In all calculations presented above, viscous 
heating only determined the thermal structure of the disk. This is sufficient in the inner 
parts of the disk at the beginning of the simulations, but leads to too low temperatures in the outer 
parts of the disk and in the later phases of the disk evolution, when the rate of gas flow 
through the disk becomes low and the irradiation is dominant. This effect leads to a different 
temperature gradient, which is important for the strength of the torques \citep{lyrapaardekooper2010}. The increased heating in 
the outer parts makes the temperature profile less steep throughout the 
disk. The temperature is still around $20\ \rm K$ at $50\ \rm AU$ and decreases with distance, while in the nonirradiated case the temperature dropped to the background 
temperature of $10\ \rm K$ at $20-30\ \rm AU$ and became nearly constant. The
temperature structure in the whole disk is set only
by the irradiation when the disk is nearly gone and viscous heating is 
unimportant for the temperature structure of the entire disk. The different profile also affects the shape of  
the convergence regions. The parts of inward migration in the disk become 
smaller with time and vanish after a few million years, but still before 
the disk dissolves \citep{KretkeLin2012}. This affects first the outer and then the inner CZ. Thus, 
the CZ is no longer a stopping point for type I migration throughout the 
complete lifetime of a disk.  
Therefore the confinement of planets in a small radius (see Appendix \ref{ap:satevo}) even when only one 
value of the photoevaporation rate is used is not the case for irradiated disks.

\begin{figure} 
        \centering \includegraphics[width=1.0\linewidth]{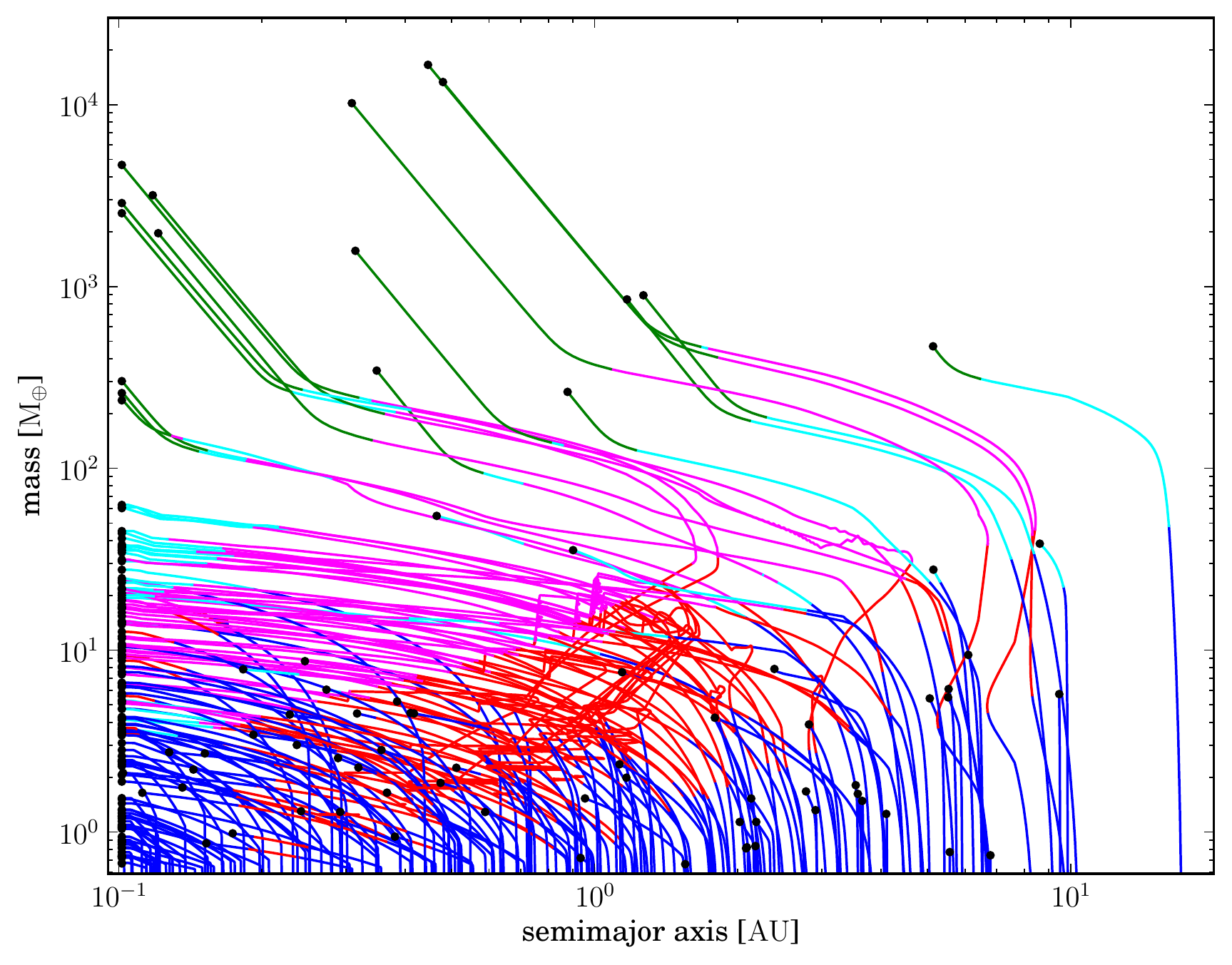} 
	\caption{Formation tracks for 250 planets of a 
          synthesis with an irradiated disk model. The meaning of the colors 
          is the same as in Fig. \ref{figsyntracks}.} 
	\label{figirradtracks} 
\end{figure} 
 
\begin{figure} 
        \centering \includegraphics[width=1.0\linewidth]{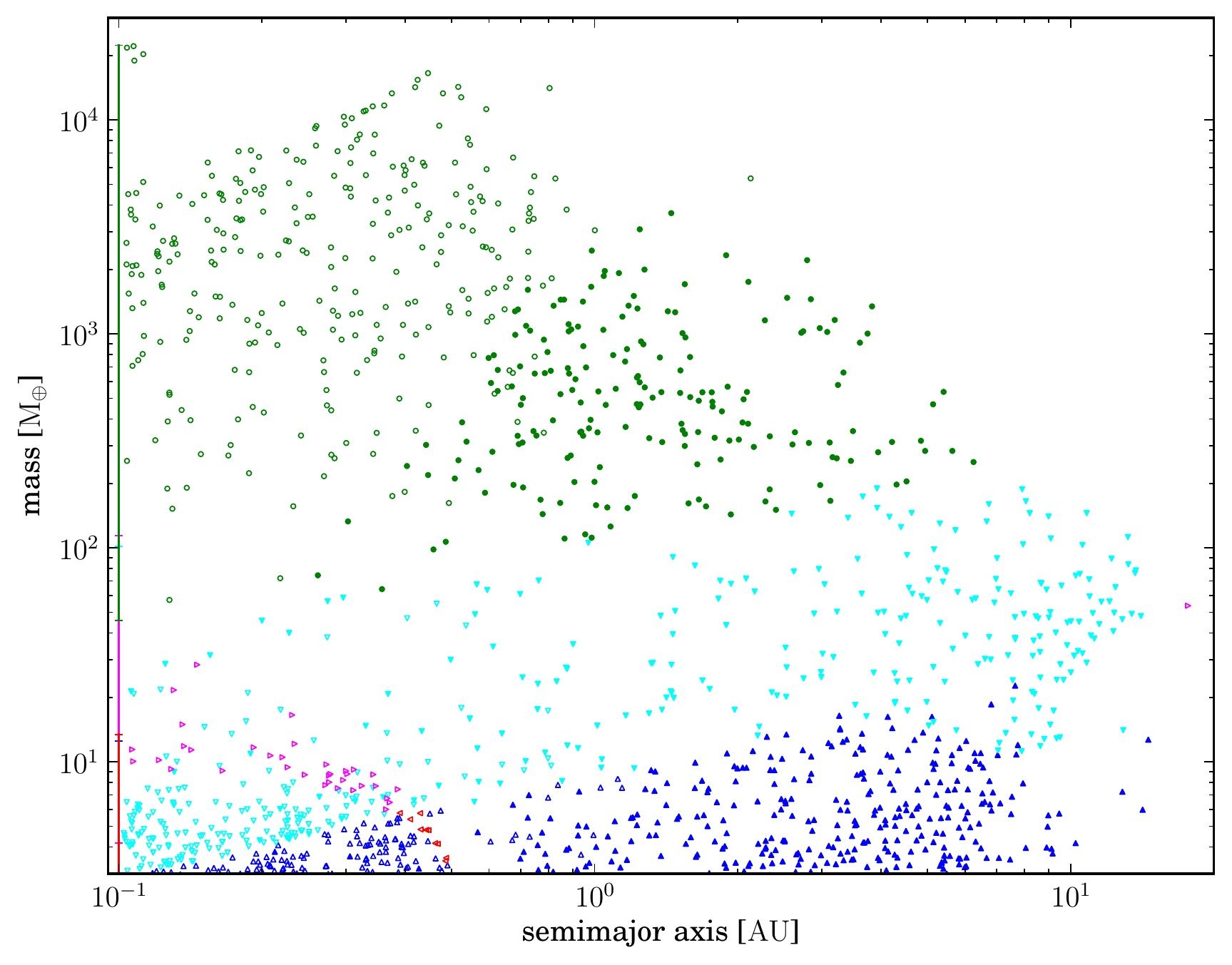} 
	\caption{Final position of the planets in the semimajor axis mass 
          diagram in the disks including stellar irradiation.  The bars at 0.1 AU indicate the mass range of the ``hot'' 
          planets in the different regimes. The meaning of the colors 
          is the same as in Fig. \ref{figsyntracks}. 
          Filled (empty) symbols represent planets that never (sometime) migrated in their 
          evolution in an adiabatic migration regime.} 
	\label{figirradAM} 
\end{figure}

Figure \ref{figirradtracks} shows the formation tracks of 250 initial conditions and Figure \ref{figirradAM} shows the final positions of the planets in the 
distance-mass diagram. In both figures color shows the different 
migration regimes a planet is in in the same ways as in Figure \ref{figsyntracks}.   
The different symbols in Figure 
\ref{figirradAM} indicate if a planet migrated in adiabatic regime 
during its evolution (empty triangles or circle) or not (filled triangles 
or circle). The new temperature structure and increase in temperature results in a 
switch into the adiabatic migration regime at a higher mass than in the nonirradiated disk calculations. Moreovre, as seen in these figures, almost 
all small planets end in a locally isothermal migration regime even when they 
went into the adiabatic regime for some part of their evolution (as shown by 
the red part of the tracks in Figure \ref{figirradtracks}) and the unfilled cyan 
triangles in Fig. \ref{figirradAM}, they end in a locally isothermal regime 
again when the disk is almost gone. 
The higher temperature than in the one in nonirradiated disks, especially at the end when almost all gas is gone and 
only minor viscous heating occurs, leads to a shorter cooling time and therefore to the transition from adiabatic 
migration into locally isothermal migration.  
 
The irradiation also leads to higher disk accretion rates in the outer regions
of the disk, since $\nu = \alpha \cs²/ \Omega \propto T$, and therefore to the
faster depletion of the outer disk  
regions. The surface density is therefore lower in the 
irradiation case than in the nonirradiated at same time of the
simulation. This leads to lower migration  
rates even in locally isothermal migration.  This compensates to some degree  
the larger inward migration zones in our disks and the transition into locally 
isothermal migration at the end of the disk lifetime. 
 
We used the same initial conditions as in all other syntheses. The different 
disk model leads to different disk lifetimes, therefore the results are only 
partially comparable. We considered 7700 planets more massive than $0.6\ 
\mearth$. From these planets $68.8\%$ migrated to the inner border of our 
computational domain. From the planets with a final distance larger than 
0.1 AU from the star $\approx 20\%$ are more massive than $100\ \mearth$, 
which is similar to the ratio in our reference synthesis.  
The larger amount of ``hot'' planets is the result from the disappearing CZs at 
toward end of the disk evolution.  

\section{Summary and conclusions} 

We have compiled a prescription for type I migration based on the latest
hydrodynamic simulations of planet-disk interactions. We tested the
influence of the prescription on the outcome of the planetary population
synthesis calculations. Our migration model is based on the combination of
results from \citet{paardekooperbaruteau2010} and different time scales to
distinguish between different migration regimes. These time scales reflect the
thermodynamical behavior of the interaction between the planet and the
surrounding disk like the horseshoe drag.  We first compared the
migration torques of this model with a model of \cite{paardekooper2011} and
with radiative-hydrodynamical simulations from \citet{kleybitsch2009} and
\citet{bitschkley2011}. The comparison of the theoretical torque curves with
data from  \citet{kleybitsch2009} and \citet{bitschkley2011} suggests that with
an adjustment of the viscous time scale in the calculation of the saturation
mass (the mass when the corotation torque starts to vanish) by a factor of
$\fvisc \approx 0.55$ our model reproduces the torque better. 

We also showed the global effects of different parameters of various migration
models in a number of population synthesis calculations. 
Here, the comparison of our nominal BMF model ($\fvisc \approx 0.55$) and the 
 migration model of \citet{paardekooper2011} indicates 
 similar results (Sect. \ref{satmasssyn}) even with the difference shown in the torque curves. 

Like \citet{lyrapaardekooper2010},\citet{KretkeLin2012}, and 
\citet{HellaryNelson2012}, we also find with our prescription that
nonisothermal type I migration leads to  convergence zones (CZ), that is, points
in a disk to which planets migrate to from the inside and outside. As in
previous studies \citep{HellaryNelson2012}, we find that the  
convergence zones move inward as the disk evolves and take the captured planets 
with it. This occurs on the slower time scale of disk evolution and therefore 
the captured planets are trapped in it and also only migrate on this time scale. 
The planets leave this zone when their horseshoe drag saturates. 
 
The difference between the migration rate of planets captured in a CZ and 
the migration rate in the saturated regime is significant. This means that an increase 
of the mass where saturation occurs by a modest factor of 2, for example,
 the time spent in rapid-saturated type I migration is significantly 
 shortened and therefore also the extent of migration. The level of saturation at a given mass, and 
 the mass at which saturation begins, are among the critical aspects for the 
 evolution of a giant planet since small changes by a factor of 2 in $\fvisc$ can change 
 the final distribution in the mass semimajor axis diagram by a measurable
 degree. But a similar degree of change in the final semimajor axis-mass 
 distribution was seen when we changed the description of the Lindblad torque 
 to the formula found by  \citet{massetcasoli2010}.   

Finally, we determined the formation tracks of a planet, illustrating that under  
certain conditions, a planet can loose almost all of its gas mass during its 
evolution. This mass loss can lead to a stop of migration for a few 
$10^5$ years when the mass loss is strong enough to desaturate the horseshoe 
drag. The reason for the envelope mass loss is a jump in solid accretion rate, 
which is caused by the migration of the planet from a solid-depleted into a 
solid-rich region of the disk. This behavior strongly depends on the 
treatment of the planetesimal disk. Here we did not evolve the disk or changed 
the random velocities because of a giant planet. Both change the 
accretion rate and therefore whether or how this mass loss occurs.

Neither did we consider random variations in the torque due to turbulent 
density variations in the disk. Recent studies showed that in some cases random migration due to turbulence 
can dominate the migration behavior for low-mass planets \citep{PapaloizouNelson2004,UribeKlahr2011}. Depending on the strength of the 
random torque, it could push planets from the inner into the outer convergence 
zone even for the stronger Lindblad torque of \citet{paardekooper2011}.  

We made no detailed comparison with observed extrasolar planet populations because the migration model 
is only a small part of the improvements to the overall model, and defer  
such comparisons to future work. But we quantified the impact of different model settings by studying the fractional yield of different planet types in a synthesis.  
 
With the new nonisothermal migration model described here or that of 
\citet{paardekooper2011} there are still about $50\%$ of all low-mass planets 
lost in the innermost part of the disk. However, 
it is about a factor two better than the isothermal migration model without 
any artificial factors. One way to reduce the number of ``lost'' planets is an 
increase of the saturation mass.  
A lower critical mass for runaway gas accretion can also help due to the 
faster transition into type II migration \citep{HoriIkoma2011}.  
Furthermore, there are hints that there are additional effects leading to outward migration of planets 
in 3D simulations of magneto-hydrodynamic disks \citep{UribeKlahr2011}.  

We also note that up to now there is a shortcoming of all analytic torque predictions: 
they all neglect the fact that the horseshoe region is over-wide compared with the prediction used 
here when $q/h^3 \approx 1$ (\citet{massetdangelo2006}). The fast growth of the width of the horseshoe 
region in that mass range (a few ten Earth masses to a hundred Earth masses) yields a boost of 
the corotation torque, which is a strong effect. In the same mass range gap opening 
and the transition into type II migration occur. And as shown in Section \ref{compmodel}, saturation 
plays a main role in the change of the direction of the torque in that mass
range as well. Moreover, our fit 
of the torques data from  \citet{kleybitsch2009} and \citet{bitschkley2011} depends on the progression 
of the torque curve in that mass range. This means that even when our fit produced the torque data well, a closer 
study of the torque in the mass range might uncover new effects, that may change the outcome of 
population synthesis calculations. 
 
Nevertheless, the CZ or a similar effect might explain the concentration of planets in clusters 
one finds in distance-mass diagrams of observed extrasolar planets for high- 
and low-mass planets \citep{Mayor2011}. In particular, the absence of
close-in, low-mass planets in \cite{mordasinialibert2009b} is not seen any
longer with the updated migration model.
 
Additional comparisons of our results with new data from radial velocity 
measurements and Kepler data will be important, especially when we combine the new 
migration model with new improvements of our model, i.e., the long term 
evolution of giant planets \citep{mordasini2012a} and the concurring evolution of multiple planets 
per disk \citep{idalin2010,Alibert2013}. Multiple cores in one disk might be collected into one CZ and form 
one larger core. Therefore the convergence region may function as a large feeding zone of solid 
matter onto a core captured in a convergence zone and thus enhance the solid 
accretion rate \citep{HellaryNelson2012, Sandor2011, Horn2012}. 
Moreover, in particular the low-mass planets of the inner cluster only
migrated through the inner convergence region. The outer boundary of this
region is due to the transition in the gas opacity at the iceline. This means
that all these planets moved only through the ice-free part of the
protoplanetary disk.  
We will also study the global 
effects of different viscous $\alpha$ values and deadzones on the behavior 
of the convergences zones with population synthesis calculations \citep[cf.][]{HasegawaPudritz2011}.
This brings us closer to a description of orbital migration without ad hoc efficiency factors.

\addcontentsline{toc}{chapter}{Bibliography} 
\bibliographystyle{aa}
\bibliography{minimalstdlib}

\begin{thebibliography}{66}
\expandafter\ifx\csname natexlab\endcsname\relax\def\natexlab#1{#1}\fi

\bibitem[{{Alexander} \& {Armitage}(2009)}]{alexanderarmitage2009}
{Alexander}, R.~D. \& {Armitage}, P.~J. 2009, \apj, 704, 989

\bibitem[{{Alibert} {et~al.}(2013){Alibert}, {Carron}, \&
  {Fortier}}]{Alibert2013}
{Alibert}, Y., {Carron}, F., \& {Fortier}, A. 2013, \aap, in review

\bibitem[{Alibert {et~al.}(2004)Alibert, Mordasini, \&
  Benz}]{alibertmordasini2004}
Alibert, Y., Mordasini, C., \& Benz, W. 2004, \aap, 417, L25

\bibitem[{{Alibert} {et~al.}(2011){Alibert}, {Mordasini}, \&
  {Benz}}]{alibertmordasini2011}
{Alibert}, Y., {Mordasini}, C., \& {Benz}, W. 2011, \aap, 526, A63

\bibitem[{Alibert {et~al.}(2005)Alibert, Mordasini, Benz, \&
  Winisdoerffer}]{alibertmordasini2005}
Alibert, Y., Mordasini, C., Benz, W., \& Winisdoerffer, C. 2005, \aap, 434, 343

\bibitem[{Armitage(2007)}]{Armitage2007}
Armitage, P.~J. 2007, The Astrophysical Journal, 665, 1381

\bibitem[{{Armitage} \& {Rice}(2005)}]{armitagerice2005}
{Armitage}, P.~J. \& {Rice}, W.~K.~M. 2005, STScI Symposium A Decade Of
  Extrasolar Planets Around Normal Stars

\bibitem[{{Baruteau} \& {Masset}(2008)}]{baruteaumasset2008}
{Baruteau}, C. \& {Masset}, F. 2008, \apj, 672, 1054

\bibitem[{{Batalha} {et~al.}(2011){Batalha}, {Borucki}, {Bryson}, {Buchhave},
  {Caldwell}, {Christensen-Dalsgaard}, {Ciardi}, {Dunham}, {Fressin},
  {Gautier}, {Gilliland}, {Haas}, {Howell}, {Jenkins}, {Kjeldsen}, {Koch},
  {Latham}, {Lissauer}, {Marcy}, {Rowe}, {Sasselov}, {Seager}, {Steffen},
  {Torres}, {Basri}, {Brown}, {Charbonneau}, {Christiansen}, {Clarke},
  {Cochran}, {Dupree}, {Fabrycky}, {Fischer}, {Ford}, {Fortney}, {Girouard},
  {Holman}, {Johnson}, {Isaacson}, {Klaus}, {Machalek}, {Moorehead},
  {Morehead}, {Ragozzine}, {Tenenbaum}, {Twicken}, {Quinn}, {VanCleve},
  {Walkowicz}, {Welsh}, {Devore}, \& {Gould}}]{batalha2011}
{Batalha}, N.~M., {Borucki}, W.~J., {Bryson}, S.~T., {et~al.} 2011, \apj, 729,
  27

\bibitem[{{Bell} \& {Lin}(1994)}]{belllin1994}
{Bell}, K.~R. \& {Lin}, D.~N.~C. 1994, \apj, 427, 987

\bibitem[{{Bitsch} \& {Kley}(2011)}]{bitschkley2011}
{Bitsch}, B. \& {Kley}, W. 2011, \aap, 536, A77

\bibitem[{{Bodenheimer} {et~al.}(2013){Bodenheimer}, {D'Angelo}, {Lissauer},
  {Fortney}, \& {Saumon}}]{bodenheimer2013}
{Bodenheimer}, P., {D'Angelo}, G., {Lissauer}, J.~J., {Fortney}, J.~J., \&
  {Saumon}, D. 2013, \apj, 770, 120

\bibitem[{{Casoli} \& {Masset}(2009)}]{casolimasset2009}
{Casoli}, J. \& {Masset}, F.~S. 2009, \apj, 703, 845

\bibitem[{{Crida} \& {Morbidelli}(2007)}]{cridamorbidelli2007}
{Crida}, A. \& {Morbidelli}, A. 2007, \mnras, 377, 1324

\bibitem[{{Crida} {et~al.}(2007){Crida}, {Morbidelli}, \&
  {Masset}}]{cridamorbidelli2006}
{Crida}, A., {Morbidelli}, A., \& {Masset}, F. 2007, \aap, 461, 1173

\bibitem[{{D'Angelo} {et~al.}(2002){D'Angelo}, {Henning}, \&
  {Kley}}]{DangeloHenning2002}
{D'Angelo}, G., {Henning}, T., \& {Kley}, W. 2002, \aap, 385, 647

\bibitem[{{Fortier} {et~al.}(2013){Fortier}, {Alibert}, {Carron}, {Benz}, \&
  {Dittkrist}}]{Fortier2013}
{Fortier}, A., {Alibert}, Y., {Carron}, F., {Benz}, W., \& {Dittkrist}, K.-M.
  2013, \aap, 549, A44

\bibitem[{Fortier {et~al.}(2007)Fortier, Benvenuto, \&
  Brunini}]{fortierbenvenuto2007}
Fortier, A., Benvenuto, O.~G., \& Brunini, A. 2007, \aap, 473, 311

\bibitem[{{Fouchet} {et~al.}(2012){Fouchet}, {Alibert}, {Mordasini}, \&
  {Benz}}]{fouchetalibert2012}
{Fouchet}, L., {Alibert}, Y., {Mordasini}, C., \& {Benz}, W. 2012, \aap, 540,
  A107

\bibitem[{{Goldreich} \& {Tremaine}(1980)}]{goldreichetremaine1980}
{Goldreich}, P. \& {Tremaine}, S. 1980, \apj, 241, 425

\bibitem[{{Guilet} {et~al.}(2013){Guilet}, {Baruteau}, \&
  {Papaloizou}}]{GuiletBaruteau2013}
{Guilet}, J., {Baruteau}, C., \& {Papaloizou}, J.~C.~B. 2013, \mnras, 430, 1764

\bibitem[{{Hasegawa} \& {Pudritz}(2011)}]{HasegawaPudritz2011}
{Hasegawa}, Y. \& {Pudritz}, R.~E. 2011, \mnras, 417, 1236

\bibitem[{{Hellary} \& {Nelson}(2012)}]{HellaryNelson2012}
{Hellary}, P. \& {Nelson}, R.~P. 2012, \mnras, 419, 2737

\bibitem[{{Hori} \& {Ikoma}(2011)}]{HoriIkoma2011}
{Hori}, Y. \& {Ikoma}, M. 2011, \mnras, 416, 1419

\bibitem[{{Horn} {et~al.}(2012){Horn}, {Lyra}, {Mac Low}, \&
  {S{\'a}ndor}}]{Horn2012}
{Horn}, B., {Lyra}, W., {Mac Low}, M.-M., \& {S{\'a}ndor}, Z. 2012, \apj, 750,
  34

\bibitem[{Ida \& Lin(2004)}]{idalin2004}
Ida, S. \& Lin, D. N.~C. 2004, \apj, 604, 388

\bibitem[{Ida \& Lin(2008)}]{idalin2008}
Ida, S. \& Lin, D. N.~C. 2008, \apj, 673, 487

\bibitem[{{Ida} \& {Lin}(2010)}]{idalin2010}
{Ida}, S. \& {Lin}, D.~N.~C. 2010, \apj, 719, 810

\bibitem[{{Kley} {et~al.}(2009){Kley}, {Bitsch}, \& {Klahr}}]{kleybitsch2009}
{Kley}, W., {Bitsch}, B., \& {Klahr}, H. 2009, \aap, 506, 971

\bibitem[{{Kley} \& {Dirksen}(2006)}]{KleyDirksen2006}
{Kley}, W. \& {Dirksen}, G. 2006, \aap, 447, 369

\bibitem[{{Kretke} \& {Lin}(2012)}]{KretkeLin2012}
{Kretke}, K.~A. \& {Lin}, D.~N.~C. 2012, \apj, 755, 74

\bibitem[{{Lubow} {et~al.}(1999){Lubow}, {Seibert}, \&
  {Artymowicz}}]{Lubow1999}
{Lubow}, S.~H., {Seibert}, M., \& {Artymowicz}, P. 1999, \apj, 526, 1001

\bibitem[{Lynden-Bell \& Pringle(1974)}]{lynden-bellpringle1974}
Lynden-Bell, D. \& Pringle, J.~E. 1974, \mnras, 168, 603

\bibitem[{{Lyra} {et~al.}(2010){Lyra}, {Paardekooper}, \& {Mac
  Low}}]{lyrapaardekooper2010}
{Lyra}, W., {Paardekooper}, S., \& {Mac Low}, M. 2010, \apjl, 715, L68

\bibitem[{{Marois} {et~al.}(2008){Marois}, {Macintosh}, {Barman}, {Zuckerman},
  {Song}, {Patience}, {Lafreni{\`e}re}, \& {Doyon}}]{Maroisetal2008}
{Marois}, C., {Macintosh}, B., {Barman}, T., {et~al.} 2008, Science, 322, 1348

\bibitem[{{Masset}(2002)}]{masset2002}
{Masset}, F.~S. 2002, \aap, 387, 605

\bibitem[{{Masset} \& {Casoli}(2009)}]{massetcasoli2009}
{Masset}, F.~S. \& {Casoli}, J. 2009, \apj, 703, 857

\bibitem[{{Masset} \& {Casoli}(2010)}]{massetcasoli2010}
{Masset}, F.~S. \& {Casoli}, J. 2010, \apj, 723, 1393

\bibitem[{{Masset} {et~al.}(2006){Masset}, {D'Angelo}, \&
  {Kley}}]{massetdangelo2006}
{Masset}, F.~S., {D'Angelo}, G., \& {Kley}, W. 2006, \apj, 652, 730

\bibitem[{{Mayor} {et~al.}(2011){Mayor}, {Marmier}, {Lovis}, {Udry},
  {S{\'e}gransan}, {Pepe}, {Benz}, {Bertaux}, {Bouchy}, {Dumusque}, {Lo Curto},
  {Mordasini}, {Queloz}, \& {Santos}}]{Mayor2011}
{Mayor}, M., {Marmier}, M., {Lovis}, C., {et~al.} 2011, ArXiv e-prints,
  1109.2497

\bibitem[{Miguel \& Brunini(2008)}]{Miguelbrunini2008}
Miguel, Y. \& Brunini, A. 2008, \mnras, 387, 463

\bibitem[{Mizuno {et~al.}(1978)Mizuno, Nakazawa, \&
  Hayashi}]{mizunonakazawa1978}
Mizuno, H., Nakazawa, K., \& Hayashi, C. 1978, Progress of Theoretical Physics,
  60, 699

\bibitem[{Mordasini {et~al.}(2009{\natexlab{a}})Mordasini, Alibert, \&
  Benz}]{mordasinialibert2009a}
Mordasini, C., Alibert, Y., \& Benz, W. 2009{\natexlab{a}}, \aap, 501, 1139

\bibitem[{Mordasini {et~al.}(2009{\natexlab{b}})Mordasini, Alibert, Benz, \&
  Naef}]{mordasinialibert2009b}
Mordasini, C., Alibert, Y., Benz, W., \& Naef, D. 2009{\natexlab{b}}, \aap,
  501, 1161

\bibitem[{{Mordasini} {et~al.}(2012{\natexlab{a}}){Mordasini}, {Alibert},
  {Georgy}, {Dittkrist}, {Klahr}, \& {Henning}}]{mordasini2012b}
{Mordasini}, C., {Alibert}, Y., {Georgy}, C., {et~al.} 2012{\natexlab{a}},
  \aap, 547, A112

\bibitem[{{Mordasini} {et~al.}(2012{\natexlab{b}}){Mordasini}, {Alibert},
  {Klahr}, \& {Henning}}]{mordasini2012a}
{Mordasini}, C., {Alibert}, Y., {Klahr}, H., \& {Henning}, T.
  2012{\natexlab{b}}, \aap, 547, A111

\bibitem[{{Mordasini} {et~al.}(2011){Mordasini}, {Dittkrist}, {Alibert},
  {Klahr}, {Benz}, \& {Henning}}]{mordasini2011}
{Mordasini}, C., {Dittkrist}, K.-M., {Alibert}, Y., {et~al.} 2011, in IAU
  Symposium, Vol. 276, IAU Symposium, ed. A.~{Sozzetti}, M.~G. {Lattanzi}, \&
  A.~P. {Boss}, 72--75

\bibitem[{{Mordasini} {et~al.}(2010){Mordasini}, {Klahr}, {Alibert}, {Benz}, \&
  {Dittkrist}}]{mordasiniklahr2010}
{Mordasini}, C., {Klahr}, H., {Alibert}, Y., {Benz}, W., \& {Dittkrist}, K.-M.
  2010, in Circumstellar Disks and Planets: Science Cases for the Second
  Generation VLTI Instrumentation, ed. S.~{Wolf}, [arXiv:1012.5281]

\bibitem[{{Nagasawa} {et~al.}(2008){Nagasawa}, {Ida}, \&
  {Bessho}}]{NagasawaIda2008}
{Nagasawa}, M., {Ida}, S., \& {Bessho}, T. 2008, \apj, 678, 498

\bibitem[{{Paardekooper} {et~al.}(2010){Paardekooper}, {Baruteau}, {Crida}, \&
  {Kley}}]{paardekooperbaruteau2010}
{Paardekooper}, S., {Baruteau}, C., {Crida}, A., \& {Kley}, W. 2010, \mnras,
  401, 1950

\bibitem[{{Paardekooper} {et~al.}(2011){Paardekooper}, {Baruteau}, \&
  {Kley}}]{paardekooper2011}
{Paardekooper}, S.-J., {Baruteau}, C., \& {Kley}, W. 2011, \mnras, 410, 293

\bibitem[{{Paardekooper} \& {Mellema}(2006)}]{PaardekooperMellema2006}
{Paardekooper}, S.-J. \& {Mellema}, G. 2006, \aap, 459, L17

\bibitem[{{Paardekooper} \& {Mellema}(2008)}]{paardekoopermellema2008}
{Paardekooper}, S.-J. \& {Mellema}, G. 2008, \aap, 478, 245

\bibitem[{Papaloizou {et~al.}(2004)Papaloizou, Nelson, \&
  Snellgrove}]{PapaloizouNelson2004}
Papaloizou, J. C.~B., Nelson, R.~P., \& Snellgrove, M.~D. 2004, \mnras, 350,
  829

\bibitem[{Papaloizou \& Terquem(1999)}]{papaloizouterquem1999}
Papaloizou, J. C.~B. \& Terquem, C. 1999, \apj, 521, 823

\bibitem[{Perri \& Cameron(1974)}]{perricameron1974}
Perri, F. \& Cameron, A. G.~W. 1974, Icarus, 22, 416

\bibitem[{Pollack {et~al.}(1996)Pollack, Hubickyj, Bodenheimer, Lissauer,
  Podolak, \& Greenzweig}]{pollackhubickyj1996}
Pollack, J.~B., Hubickyj, O., Bodenheimer, P., {et~al.} 1996, Icarus, 124, 62

\bibitem[{{Rasio} \& {Ford}(1996)}]{RasioFord1996}
{Rasio}, F.~A. \& {Ford}, E.~B. 1996, Science, 274, 954

\bibitem[{{S{\'a}ndor} {et~al.}(2011){S{\'a}ndor}, {Lyra}, \&
  {Dullemond}}]{Sandor2011}
{S{\'a}ndor}, Z., {Lyra}, W., \& {Dullemond}, C.~P. 2011, \apjl, 728, L9

\bibitem[{{Saumon} {et~al.}(1995){Saumon}, {Chabrier}, \& {van
  Horn}}]{SaumonChabrier1995}
{Saumon}, D., {Chabrier}, G., \& {van Horn}, H.~M. 1995, \apjs, 99, 713

\bibitem[{Shakura \& Sunyaev(1973)}]{shakurasunayev1973}
Shakura, N.~I. \& Sunyaev, R.~A. 1973, Astron. Astrophys., 24, 337

\bibitem[{Tanaka {et~al.}(2002)Tanaka, Takeuchi, \& Ward}]{tanakatakeuchi2002}
Tanaka, H., Takeuchi, T., \& Ward, W.~R. 2002, \apj, 565, 1257

\bibitem[{Thommes {et~al.}(2008)Thommes, Matsumura, \& Rasio}]{Thommesetal2008}
Thommes, E.~W., Matsumura, S., \& Rasio, F.~A. 2008, Science, 321, 814

\bibitem[{{Uribe} {et~al.}(2011){Uribe}, {Klahr}, {Flock}, \&
  {Henning}}]{UribeKlahr2011}
{Uribe}, A.~L., {Klahr}, H., {Flock}, M., \& {Henning}, T. 2011, \apj, 736, 85

\bibitem[{Veras \& Armitage(2004)}]{verasarmitage2004}
Veras, D. \& Armitage, P.~J. 2004, \mnras, 347, 613

\bibitem[{{Yamada} \& {Inaba}(2012)}]{YamadaInaba2012}
{Yamada}, K. \& {Inaba}, S. 2012, \mnras, 424, 2746

\end{thebibliography}

\begin{appendix}
\section{Saturation and disk evolution} 
\label{ap:satevo} 
\begin{figure} 
        \centering \includegraphics[width=1.0\linewidth]{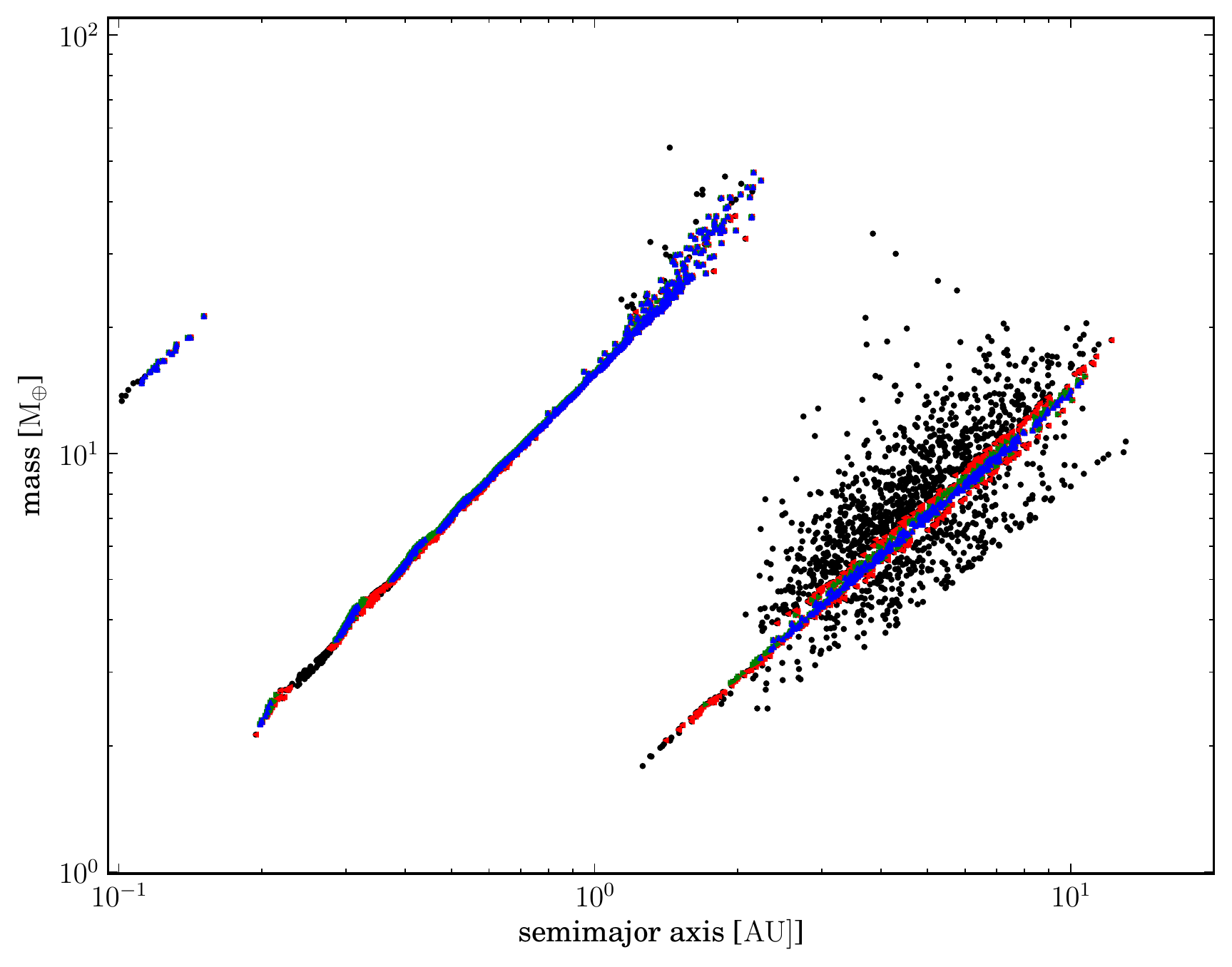} 
	\caption{Position of a planet in the nominal synthesis calculation 
          when it transitions from the unsaturated adiabatic migration regime 
          into the saturated adiabatic migration regime. Colored are the 
          planet which distance is less than 3\% (blue, up facing triangles), 
          5\% (green, right facing triangles) or 10\% (red, left facing triangles) from a CZ. } 
	\label{figsatpoint} 
\end{figure} 
 
From the positions of planets at the time of saturation in the 
formation tracks in Figure \ref{figsyntracks} in Section \ref{refsyncalc} one sees that most positions lie on two 
lines, one for the planets of the inner and one for the outer CZ. We study here the reasons for this feature.

 Figure \ref{figsatpoint} shows the positions of the planets at the transition from 
unsaturated to saturated adiabatic migration, the most 
common transition into a saturated migration regime. The positions of all 
planets undergoing this transition are shown in black, while the colored points 
show that those with a distance between their semimajor axis and the position of 
the CZ at that time are fewer than 10\% of the semimajor 
axis. There are two large groups, again one of the inner and one of the outer 
convergence region. Almost all planets of the inner 
group are in the CZ when they saturate, while the outer ones are much more 
spread out. But the planets that are in the CZ form a line here as well. There is 
a third CZ inside of 0.3 AU in the most massive disks in the 
synthesis, which leads to a third minor group. However, this CZ evolves quickly and 
disappears after the first few 0.01 Myr and all associated planets end at 
0.1 AU. The dozen points inside of 0.2 AU in Fig. \ref{figsatpoint} correspond to planets in this 
small CZ. 
 
This behavior results from the interaction of the following points: 
 
One can calculate the saturation mass as a function of the orbital distance $a$ 
and time $t$ by setting $s_2 = 1$ in Eq. \ref{s2fvisc}. There, $\fvisc$ 
 is 1, 0.55 or 0.125 in each of the model versions.  
\beq 
\msat(a,t) = \frac{h(a,t)}{1.16} \left(\frac{8\pi \mstar \nu(a,t)}{3 \fvisc}\right)^{\frac{2}{3}} \frac{1}{(G a)^{1/3}}. \label{eqmsat} 
\eeq 
For a fixed orbital distance and stellar mass it only depends on the disk 
aspect ratio $h$ and the viscosity $\nu$. Both quantities are decreasing with 
time, as the disk mass decreases, and therefore, the saturation mass also becomes smaller as the disk evolves. 
 
While the photoevaporation rate is important for 
the lifetime of a disk, the constant value of $\alpha$ in all simulations of one synthesis results 
in a similar disk structure in the part of the disk where viscosity is 
dominant. There, the disks go through the same series of 
disk states (radial profile of temperature, surface density, etc.) and only the speed with which the disks go through the states is 
different and depends on the photoevaporation rate. Given one semimajor axis
$a$, there is only one disk state where the inner (our outer) CZ lie at this
position. Therefore also $h$ and $\nu$ are fixed for this semimajor axis of
the CZ. Therefore, the semimajor axis of the CZ corresponds to only one 
saturation mass. And as the CZ moves inward while the disk evolves, the 
saturation mass decreases. Both processes approximately follow power-laws and thus 
we see a line-like structure for the planets that saturate while they are in one 
CZ. Planets that saturate early in the disk evolution do so at a higher mass 
and farther out than planets that saturate in later times of disk evolution.  
 
Finally, the spread in the outer group results from planets that saturated 
before they reached the CZ. These are planets in disks with high solid 
surface densities where the planet cores can grow fast. Compared with the inner 
group, the outer group also contains more planets that saturate outside of the CZ. 
The larger amount of solids outside of the iceline leads to higher accretion rates. 
The scatter is reduced when the saturation mass is increased 
by reducing $\fvisc$. The higher saturation mass gives the planets more time 
to migrate to the CZ and to saturate there.

\section{ Impact of numerical parameters} 
\label{minoreffects} 
We made several population syntheses calculation to test the effects of 
different numerical parameters and comment on the effects here. 
 
We made calculations with the STD model and exponents $b=2.0$ and  $b=10.0$ 
(nominal value $b=4.0$) in the 
transition function between locally isothermal and adiabatic migration 
regime (Eq. \ref{ct1v2}). This only has a small effect for low-mass planets, while for massive planets 
the final semimajor axis and mass is almost the same for different values 
of $b$. At smaller masses no clear pattern can be seen. An increase of $b$ can lead 
to either more or less massive planets and to either larger or smaller distances 
from the star of a few per cent.
 
We also made calculations with $b=4.0$ and a hard jump for the transition 
function between type I and type II migration (Eq. \ref{cfinal}) (nominal
value $b=10$. This only affects  massive 
planets and gives only a change in distance from the star of a few percent, 
with a larger $b$ leading to planets farther away from the star but almost no 
change in the final mass.  

Overall, the range of the different parameters studied here only leads to minor changes in the overall distribution of planets in semimajor axis and mass.

\end{appendix}
\listofobjects 
\end{document}